\def\us{\bds{\hat{s}}}

\def\siss{\sin 2i}
\def\cee{\cos 2\varepsilon}
\def\see{\sin 2\varepsilon}
\def\cis{\cos i}
\def\sis{\sin i}
\def\ce{\cos\varepsilon}
\def\se{\sin\varepsilon}

\def\Kr{\kA\bds\cdot\uR}
\def\Kl{\kA\bds\cdot\ul}
\def\Km{\kA\bds\cdot\um}
\def\Kn{\kA\bds\cdot\uN}
\def\Klp{\kp\bds\cdot\ul}
\def\Kmp{\kp\bds\cdot\um}
\def\Knp{\kp\bds\cdot\uN}

\def\ul{{\bds{\hat{l}}}}
\def\um{{\bds{\hat{m}}}}

\def\sA{{\bds{\mathcal{S}}}^\textrm{A}}
\def\sB{{\bds{\mathcal{S}}}^\textrm{B}}

\def\kA{{\bds{\hat{S}}}^\textrm{A}}
\def\kB{{\bds{\hat{S}}}^\textrm{B}}

\def\kp{{\bds{\hat{S}}}^\textrm{p}}

\def\kx{{\hat{S}}_x}
\def\ky{{\hat{S}}_y}

\def\mA{M_\textrm{A}}
\def\mB{M_\textrm{B}}
\def\uR{\boldsymbol{\hat{\rho}}}
\def\uT{\boldsymbol{\hat{\sigma}}}
\def\uN{\boldsymbol{\hat{\nu}}}
\def\uI{{\boldsymbol{\hat{e}}}_x}
\def\uJ{{\boldsymbol{\hat{e}}}_y}
\def\uK{{\boldsymbol{\hat{e}}}_z}
\def\nk{n_{\rm b}}

\def\snf{\sin f}
\def\csf{\cos f}
\def\cu{\cos u}
\def\su{\sin u}

\def\Pb{P_{\rm b}}

\def\rfr#1{Equation~(\ref{#1})}
\def\rfrs#1#2{Equations~(\ref{#1})~to~(\ref{#2})}
\def\Rfr#1{Eq. (\ref{#1})}

\def\derp#1#2{\rp{\partial{#1}}{\partial{#2}}}
\def\dert#1#2{\frac{{{\textrm{d}}}{#1}}{{{\textrm{d}}}{#2}}}

\def\virg#1{``#1"}

\def\eqi{\begin{equation}}
\def\eqf{\end{equation}}
\def\eqia{\begin{eqnarray}}
\def\eqfa{\end{eqnarray}}

\def\rp#1#2{{#1\over#2}}
\def\lb#1{\label{#1}}

\def\bds#1{\boldsymbol{#1}}

\def\co{\cos\omega}
\def\so{\sin\omega}

\def\cO{\cos\Omega}
\def\sO{\sin\Omega}

\def\cI{\cos I}
\def\sI{\sin I}

\def\ton#1{\left(#1\right)}
\def\qua#1{\left[#1\right]}
\def\grf#1{\left\{#1\right\}}
\def\ang#1{\left\langle #1\right\rangle}
\documentclass{aastex}

\usepackage{hyperref}
\usepackage{float}
\usepackage{amsmath,textgreek,w-greek,wasysym}
\usepackage{amscd,lineno}
\usepackage{amssymb}
\usepackage{graphicx,epsfig}
\usepackage{txfonts}
\usepackage{xr-hyper}

\RequirePackage{color}

\newcommand{\emaila}{lorenzo.iorio@libero.it}

\allowdisplaybreaks[1]

\begin{document}

\title{Post-Keplerian perturbations of the orbital time shift in binary pulsars: an analytical formulation with applications to the Galactic Center}

\shortauthors{L. Iorio}

\author{Lorenzo Iorio\altaffilmark{1} }
\affil{Ministero dell'Istruzione, dell'Universit\`{a} e della Ricerca
(M.I.U.R.)-Istruzione
\\ Permanent address for correspondence: Viale Unit\`{a} di Italia 68, 70125, Bari (BA),
Italy}

\email{\emaila}

\begin{abstract}
We develop a general approach to analytically calculate the perturbations $\Delta\delta\tau_\textrm{p}$ of the orbital component of the change $\delta\tau_\textrm{p}$ of the times of arrival of the pulses emitted by a binary pulsar p induced by the post-Keplerian accelerations due to the mass quadrupole $Q_2$, and the post-Newtonian gravitoelectric (GE) and Lense-Thirring (LT) fields. We apply our results to the so-far still hypothetical scenario involving a pulsar orbiting the Supermassive Black Hole in in the Galactic Center at Sgr A$^\ast$. We also evaluate the gravitomagnetic and quadrupolar Shapiro-like propagation delays $\delta\tau_\textrm{prop}$. By assuming the orbit of the existing S2 main sequence star and a time span as long as its orbital period $\Pb$, we obtain $\left|\Delta\delta\tau_\textrm{p}^\textrm{GE}\right|\lesssim 10^3~\textrm{s},~\left|\Delta\delta\tau_\textrm{p}^\textrm{LT}\right|\lesssim 0.6~\textrm{s},\left|\Delta\delta\tau_\textrm{p}^{Q_2}\right|\lesssim 0.04~\textrm{s}$. Faster $\ton{\Pb = 5~\textrm{yr}}$ and more eccentric $\ton{e=0.97}$ orbits would imply net shifts per revolution as large as $\left|\ang{\Delta\delta\tau_\textrm{p}^\textrm{GE}}\right|\lesssim 10~\textrm{Ms},~\left|\ang{\Delta\delta\tau_\textrm{p}^\textrm{LT}}\right|\lesssim 400~\textrm{s},\left|\ang{\Delta\delta\tau_\textrm{p}^{Q_2}}\right|\lesssim 10^3~\textrm{s}$, depending on the other orbital parameters and the initial epoch. For the propagation delays, we have $\left|\delta\tau_\textrm{prop}^\textrm{LT}\right|\lesssim 0.02~\textrm{s},~\left|\delta\tau_\textrm{prop}^{Q_2}\right|\lesssim 1~\mu\textrm{s}$. The results for the mass quadrupole and the Lense-Thirring field depend, among other things, on the spatial orientation of the spin axis of the Black Hole. The expected precision in pulsar timing in Sgr A$^\ast$ is of the order of $100~\mu\textrm{s}$, or, perhaps, even $1-10~\mu\textrm{s}$. Our method is, in principle, neither limited just to some particular orbital configuration nor to the dynamical effects considered in the present study.
\end{abstract}

keywords{
gravitation--celestial mechanics--binaries: general--pulsars: general--stars: black holes
}
\section{Introduction}
In a binary hosting at least one emitting pulsar\footnote{See, e.g., \citet{Wex14,Kaspi15} and references therein.} p, the time of arrivals $\tau_\textrm{p}$ of the emitted radio pulses changes primarily because of the orbital motion about the common center of mass caused by the gravitational tug of the unseen companion c which can be, in principle, either a main sequence star or an astrophysical compact object like, e.g., another neutron star which does not emit or whose pulses are, for some reasons, not detectable, a white dwarf or, perhaps, even a black hole \citep{1999ApJ...514..388W}. Such a periodic variation $\delta\tau_\textrm{p}\ton{f}$ can be modeled as the ratio of the projection of the barycentric orbit $\mathbf{r}_\textrm{p}$ of the pulsar p onto the line of sight  to the speed of light $c$ \citep{1991PhRvL..66.2549D,2000ApJ...544..921K}. By assuming a coordinate system centered in the binary's center of mass whose reference $z$-axis points toward the observer along the line of sight in such a way that the reference $\grf{x,~y}$ plane coincides with the plane of the sky, we have
\eqi
\delta\tau_\textrm{p}\ton{f} = \rp{\mathrm{r}_z^\textrm{p}}{c}=\rp{r_\textrm{p}\sI \su}{c}=\rp{a_\textrm{p}\ton{1-e^2}\sI\su}{c\ton{1+e\csf}}=\rp{m_\textrm{c}p\sI\sin\ton{\omega + f}}{m_\textrm{tot}c\ton{1+e\csf}}.\lb{tau}
\eqf
In obtaining \rfr{tau}, which is somewhat the analogous of the range in Earth-Moon or Earth-planets studies \citep{1991PhRvL..66.2549D}, we used the fact that,
to the Keplerian level, the barycentric semimajor axis of the pulsar A is
\eqi
a_\textrm{p}\simeq\ton{\rp{m_\textrm{c}}{m_\textrm{tot}}}a.
\eqf
In a purely Keplerian scenario, there is no net variation $\ang{\Delta\delta\tau_\textrm{p}}$ over a full orbital cycle.

In this paper, we illustrate a relatively simple and straightforward approach to analytically calculate the impact that several post-Keplerian (pK) features of motion, both Newtonian (quadrupole) and post-Newtonian (1pN static and stationary fields), have on such a key observable. As such, we will analytically calculate the corresponding net time delays per revolution $\ang{\Delta\delta\tau_\textrm{p}}$; the instantaneous shifts $\Delta\delta\tau_\textrm{p}\ton{f}$ will be considered as well in order to cope  with systems exhibiting very long orbital periods with respect to the time spans usually adopted for data collection. Our strategy has a general validity since, in principle, it can be extended to a wide range of dynamical effects, irrespectively of their physical origin, which may  include, e.g., modified models of gravity as well. Furthermore, it is applicable to systems whose constituents may have arbitrary masses and orientations of their spin axes, and orbital configurations.
Thus, more realistic sensitivity analyses, aimed to both re-interpreting already performed studies and designing future targeted ones, could be conducted in view of a closer correspondence with which is actually measured. We we will also take into account the Shapiro-like time delays due to the propagation of the electromagnetic waves emitted by the visible pulsar(s) throughout the spacetime deformed by axisymmetric departures from spherical symmetry of the deflecting bodies  \citep{DamDer86, 1991SvA....35..523K, 1992PhRvD..45.1840D, 1995MNRAS.274.1029D, 1997JMP....38.2587K, 1999ApJ...514..388W, 2003AJ....125.1580K, 2011CQGra..28a5009Z}.

Our results, which are not intended to replace dedicated, covariance-based real data analyses, being, instead, possible complementary companions,  will be applied to the so far putative scenario involving emitting radiopulsars, not yet detected, orbiting the Supermassive Black Hole (SMBH) in the Galactic Center (GC) at Sgr A$^\ast$ \citep{2010ApJ...720.1303A,2013IAUS..291..382E,2014ApJ...784..106Z,2016IJMPD..2530029K,2016CQGra..33k3001J,2016ApJ...818..121P,2017IJMPD..2630001G}. Moreover, we will perform also quantitative sensitivity analyses on the measurability of frame-dragging and quadrupolar-induced time delays in such a hypothesized system. In principle, our results may be applicable even to anthropogenic binaries like, e.g., those contrived in past concept studies to perform tests of fundamental physics in space \citep{1988AJ.....95..576N,2008IJMPD..17..453S}, or continuously emitting transponders placed on the surface of some moons of larger astronomical bodies.

The paper is organized as follows. Section \ref{metodo} details the calculational approach. The 1pN Schwarzschild-type gravitoelectric effects are calculated in Section \ref{GEeffects}, while Section \ref{LTeffects} deals with the 1pN gravitomagnetic ones.  The impact of the quadrupole mass moment of the SMBH is treated in Section \ref{Qeffects}.  Section \ref{fine} summarizes our findings.
\section{Outline of the proposed method}\lb{metodo}
If the motion of a binary is affected by some relatively small post-Keplerian (pK) acceleration $\bds A$, either Newtonian or post-Newtonian (pN) in nature,  its impact on the projection of the orbit onto the line of sight can be calculated perturbatively as follows.
\citet{1993CeMDA..55..209C} analytically worked out the instantaneous changes of the radial, transverse and out-of-plane components
$\mathrm{r}_{\rho},~\mathrm{r}_{\sigma},~\mathrm{r}_{\nu}$  of the position  vector $\mathbf{r}$, respectively, for the relative motion of a test particle about its primary: they are
\begin{align}
\Delta \mathrm{r}_{\rho}\ton{f} \lb{rR}&= \rp{r\ton{f}}{a}\Delta a\ton{f} -a\cos f\Delta e\ton{f} +\rp{ae\sin f}{\sqrt{1-e^2}}\Delta\mathcal{M}\ton{f}, \\ \nonumber \\
\Delta \mathrm{r}_{\sigma}\ton{f} \lb{rT}&= a\sin f\qua{1 + \rp{r\ton{f}}{p}}\Delta e\ton{f} + r\ton{f}\qua{\cI\Delta\Omega\ton{f}+\Delta\omega\ton{f}} +\rp{a^2}{r\ton{f}}\sqrt{1-e^2}\Delta\mathcal{M}\ton{f}, \\ \nonumber \\
\Delta \mathrm{r}_{\nu}\ton{f} \lb{rN}&= r\ton{f}\qua{\sin u~\Delta I\ton{f} -\sI\cos u~\Delta\Omega\ton{f}}.
\end{align}
In \rfrs{rR}{rN}, the instantaneous changes $\Delta a\ton{f},~\Delta e\ton{f},~\Delta I\ton{f},~\Delta\Omega\ton{f},~\Delta\omega\ton{f}$  are to be calculated as
\eqi
\Delta\kappa\ton{f}=\int_{f_0}^f\dert{\kappa}{t} \dert{t}{f^{'}} df^{'},~\kappa=a,~e,~I,~\Omega,~\omega,\lb{Dk}
\eqf
where the time derivatives $d\kappa/dt$ of the Keplerian orbital elements $\kappa$ are to be taken from the right-hand-sides of the Gauss equations
\begin{align}
\dert a t \lb{dadt}& = \rp{2}{\nk\sqrt{1-e^2}}\qua{e A_{\rho} \snf + A_{\sigma}\ton{\rp{p}{r}}}, \\ \nonumber \\
\dert e t \lb{dedt}& = \rp{\sqrt{1-e^2}}{\nk a}\grf{A_{\rho} \snf + A_{\sigma}\qua{\csf + \rp{1}{e}\ton{1 - \rp{r}{a}} } }, \\ \nonumber \\
\dert I t \lb{dIdt}& = \rp{1}{\nk a \sqrt{1 - e^2}}A_{\nu}\ton{\rp{r}{a}}\cu, \\ \nonumber \\
\dert \Omega t \lb{dOdt}& = \rp{1}{\nk a \sI\sqrt{1 - e^2}}A_{\nu}\ton{\rp{r}{a}}\su, \\ \nonumber \\
\dert \omega t \lb{dodt}& = -\cI\dert\Omega t + \rp{\sqrt{1-e^2}}{\nk a e}\qua{ -A_{\rho}\csf + A_{\sigma}\ton{1 + \rp{r}{p}}\snf },
\end{align}
evaluated onto the  Keplerian ellipse given by
\eqi
r=\rp{p}{1+e\cos f}\lb{Kepless}
\eqf and assumed as unperturbed reference trajectory; the same holds also for
\eqi
\dert t f = \rp{r^2}{\sqrt{\mu p}}= \rp{\ton{1-e^2}^{3/2}}{\nk\ton{1+e\cos f}^2}\lb{dtdfKep}
\eqf
entering \rfr{Dk}. The case of the mean anomaly $\mathcal{M}$ is subtler, and requires more care.
Indeed, in the most general case encompassing the possibility that the mean motion $\nk$ is time-dependent because of some physical phenomena, it can be written as\footnote{The mean anomaly at epoch is denoted as $\eta$ by \citet{Nobilibook87}, $l_0$ by \citet{1991ercm.book.....B}, and $\epsilon^{'}$ by \citet{2003ASSL..293.....B}. It is a \virg{slow} variable in the sense that its time derivative vanishes in the limit $\bds A\rightarrow 0$; cfr. with \rfr{detadt}. } \citep{Nobilibook87,1991ercm.book.....B,2003ASSL..293.....B}
\eqi
\mathcal{M}\ton{t} = \eta + \int_{t_0}^{t} \nk\ton{t^{'}}dt^{'};\lb{Mt}
\eqf
the Gauss equation for the variation of the mean anomaly at epoch is\footnote{It is connected with the Gauss equation for the variation of the time of passage at pericenter $t_p$ by $d\eta/dt = -\nk dt_p/dt$.} \citep{Nobilibook87,1991ercm.book.....B,2003ASSL..293.....B}
\eqi
\dert\eta t = - \rp{2}{\nk a}A_{\rho}\ton{\rp{r}{a}} -\rp{\ton{1-e^2}}{\nk a e}\qua{ -A_{\rho}\csf + A_{\sigma}\ton{1 + \rp{r}{p}}\snf }\lb{detadt}.
\eqf
If $\nk$ is constant, as in the Keplerian case, \rfr{Mt} reduces to the usual form
\eqi
\mathcal{M}\ton{t}= \eta + \nk\ton{t-t_0}.
\eqf
In general, when a disturbing acceleration is present, the semimajor axis $a$ does vary according to \rfr{dadt}; thus, also the mean motion $\nk$ experiences a change\footnote{We neglect the case $\mu\ton{t}$.}
\eqi
\nk\rightarrow \nk+\Delta\nk\ton{t}
\eqf
which can be calculated in terms of the true anomaly $f$ as
\eqi
\Delta \nk\ton{f}=\derp{\nk}{a}\Delta a\ton{f}= -\rp{3}{2}\rp{\nk}{a}\int_{f_0}^f\dert a t \dert{t}{f^{'}}df^{'}\lb{Dn}
\eqf
by means of \rfr{dadt} and \rfr{dtdfKep}.
Depending on the specific perturbation at hand, \rfr{Dn} does not generally vanish.
Thus, the total change experienced by the mean anomaly $\mathcal{M}$ due to the disturbing acceleration $\bds A$ can be obtained as
\eqi
\Delta\mathcal{M}\ton{f} = \Delta\eta\ton{f} + \int_{t_0}^{t}\Delta\nk\ton{t^{'}} dt^{'},\lb{anom}
\eqf
where
\begin{align}
\Delta\eta\ton{f} &= \int_{f_0}^f\dert\eta t \dert{t}{f^{'}} df^{'}, \\ \nonumber \\
\int_{t_0}^{t}\Delta\nk\ton{t^{'}} dt^{'} \lb{inte}& = -\rp{3}{2}\rp{\nk}{a}\int_{f_0}^f\Delta a\ton{f^{'}}\dert{t}{f^{'}}df^{'}.
\end{align}
In the literature, the contribution due to \rfr{inte} has been often neglected.
An alternative way to compute the perturbation of the mean anomaly with respect to \rfr{anom} implies the use of the mean longitude $\lambda$ and the longitude of pericenter $\varpi$.
It turns out that\footnote{The mean longitude at epoch is denoted as $\epsilon$ by \citet{Nobilibook87,1989racm.book.....S,1991ercm.book.....B,2003ASSL..293.....B}. It is better suited than $\eta$ at small inclinations \citep{2003ASSL..293.....B}.} \citep{1989racm.book.....S}
\eqi
\Delta\mathcal{M}\ton{f} = \Delta\epsilon\ton{f} - \Delta\varpi\ton{f} + \int_{t_0}^{t}\Delta\nk\ton{t^{'}} dt^{'},\lb{longi}
\eqf
where the Gauss equations for the variation of $\varpi,~\epsilon$ are \citep{Nobilibook87,1989racm.book.....S,1991ercm.book.....B,2003ASSL..293.....B}
\begin{align}
\dert\varpi t & =2\sin^2\ton{\rp{I}{2}}\dert\Omega t +\rp{\sqrt{1-e^2}}{\nk a e}\qua{-A_{\rho}\cos f+A_{\sigma}\ton{1+\rp{r}{p}}\sin f}, \\ \nonumber \\
\dert\epsilon t & = \rp{e^2}{1+\sqrt{1-e^2}}\dert\varpi t +2\sqrt{1-e^2}\dert\Omega t-\rp{2}{\nk a}A_{\rho}\ton{\rp{r}{a}}.
\end{align}
It must be remarked that, depending on the specific perturbing acceleration $\bds A$ at hand, the calculation of \rfr{inte} may turn out to be rather uncomfortable.

The instantaneous change experienced by the projection of the binary's relative motion onto the line of sight can be extracted from \rfrs{rR}{rN} by taking the  $z$ component $\Delta \mathrm{r}_z$ of the vector
\eqi
\Delta \mathbf{r} = \Delta \mathrm{r}_{\rho}~\uR + \Delta \mathrm{r}_{\sigma}~\uT + \Delta \mathrm{r}_{\nu}~\uN
\eqf
expressing the perturbation experienced by the binary's relative position vector $\mathbf{r}$.
It is
\begin{align}
\Delta \mathrm{r}_z\ton{f} \nonumber \lb{Dzf}& = \rp{\ton{1-e^2}\sI\su}{1+e\csf}\Delta a\ton{f} + \\ \nonumber \\
\nonumber &+a\sI\qua{\ton{1+\rp{1}{1+e\csf}}\snf\cu  -\csf\su  }\Delta e\ton{f} + \\ \nonumber \\
\nonumber & + \rp{a\ton{1-e^2}\cI\su}{1+e\csf}\Delta I\ton{f} + \rp{a\ton{1-e^2}\sI\cu}{1+e\csf}\Delta\omega\ton{f} + \\ \nonumber \\
&+\rp{a\sI\ton{e\co + \cu}}{\sqrt{1-e^2}}\Delta\mathcal{M}\ton{f}.
\end{align}
It is possible to express the true anomaly as a function of time through the mean anomaly according to \citet[p.~77]{1961mcm..book.....B}
\eqi
f\ton{t} = \mathcal{M}\ton{t} + 2\sum_{s = 1}^{s_\textrm{max}}\rp{1}{s}\grf{ J_s\ton{se} + \sum_{j = 1}^{j_\textrm{max}}\rp{\ton{1-\sqrt{1-e^2}}^j}{e^j}\qua{ J_{s-j}\ton{se} + J_{s+j}\ton{se}  }  }\sin s\mathcal{M}\ton{t}, \lb{fMt}
\eqf
where $J_k\ton{se}$ is the Bessel function of the first kind of order $k$ and $s_\textrm{max},~j_\textrm{max}$ are some values of the summation indexes $s,~j$ adequate for the desired accuracy level.
Having at disposal such analytical time series yielding the time-dependent pattern  of \rfr{Dzf} allows one to easily study some key features of it such as, e.g., its extrema along with the corresponding epochs and the values of some unknown parameters which may enter the disturbing acceleration.
The net change per orbit $\ang{\Delta \mathrm{r}_z}$ can be obtained by calculating \rfr{Dzf}
with $f=f_0+2\uppi$, and using \rfr{Dk} and \rfrs{anom}{inte} integrated from $f_0$ to $f_0+2\uppi$.

In order to have the change of the times of arrival of the pulses from the binary's pulsar p, \rfr{Dzf} and its orbit averaged expression have to be scaled by $m_\textrm{c} m_\textrm{tot}^{-1}c^{-1}$.

In the following, we will look at three pK dynamical effects: the Newtonian deviation from spherical symmetry of the binary's bodies due to their quadrupole mass moments, and the velocity-dependent 1pN static (gravitoelectric) and stationary (gravitomagnetic) accelerations responsible of the time-honored anomalous Mercury's perihelion precession and the Lense-Thirring frame-dragging, respectively.
\section{The 1pN gravitoelectric effect}\lb{GEeffects}
Let us start with the static component of the 1pN field which, in the case of our Solar System, yields the formerly anomalous perihelion precession of Mercury  of $\dot\varpi_{\mercury}=42.98~\textrm{arcsec~cty}^{-1}$ \citep{1986Natur.320...39N}.

The 1pN gravitoelectric, Schwarzschild-type, acceleration of the relative motion is, in General Relativity, \citep{1989racm.book.....S}
\eqi
{\bds A}_\textrm{GE}=\rp{\mu}{c^2 r^2}\grf{\qua{\ton{4 + 2 \xi}\rp{\mu}{r} -\ton{1 + 3\xi}\ton{\mathbf{v}\bds\cdot\mathbf{v}}  + \rp{3}{2}\xi\ton{\mathbf{v}\bds\cdot\uR}^2  }\uR +\ton{4 - 2\xi}\ton{\mathbf{v}\bds\cdot\uR}\mathbf{v} }.\lb{AGE}
\eqf
By projecting \rfr{AGE} onto the radial, transverse, out-of-plane unit vectors $\uR,~\uT,~\uN$, its corresponding components are
\begin{align}
A_{\rho}^\textrm{GE} \lb{ARGE}& =\rp{\mu^2\ton{1 + e \csf}^2 \qua{ \ton{4 - 13\xi} e^2  + 4\ton{3 - \xi} + 8\ton{1 - 2\xi}e\csf  -\ton{8 - \xi}e^2\cos 2f   } }{4 c^2 a^3\ton{1-e^2}^3 }, \\ \nonumber \\
A_{\sigma}^\textrm{GE} \lb{ATGE}& = \rp{2\mu^2 \ton{1 + e \csf}^3 \ton{2 - \xi}e\snf}{c^2 a^3\ton{1 - e^2}^3}, \\ \nonumber \\
A_{\nu}^\textrm{GE} \lb{ANGE}& = 0.
\end{align}
Here, we use the true anomaly $f$ since it turns out computationally more convenient.

The resulting net shifts per orbit of the osculating Keplerian orbital elements, obtained by integrating \rfr{Dk} and \rfrs{anom}{inte} from $f_0$ to $f_0+2\uppi$, are
\begin{align}
\ang{\Delta a^\textrm{GE}} \lb{DaeIOGE}&=\ang{\Delta e^\textrm{GE}} = \ang{\Delta I^\textrm{GE}} = \ang{\Delta\Omega^\textrm{GE}} = 0, \\ \nonumber \\
\ang{\Delta\omega^\textrm{GE}}\lb{Domega}& = \ang{\Delta\varpi^\textrm{GE}} = \rp{6\uppi\mu}{c^2 p}, \\ \nonumber \\
\ang{\Delta\mathcal{M}^\textrm{GE}} \lb{DMGE} \nonumber & = \rp{\uppi\mu}{4c^2 a\ton{1-e^2}^2}\grf{8\ton{-9+2\xi}  +4e^4\ton{-6+7\xi} +e^2\ton{-84+76\xi} + \right.\\ \nonumber\\
\nonumber &+\left.3e\qua{ 8\ton{-7+3\xi}  +e^2\ton{-24+31\xi }   }\cos f_0 + \right. \\ \nonumber \\
&+\left. 3e^2\qua{ 4\ton{-5+4\xi}\cos 2f_0  +e\xi\cos 3 f_0  }}.
\end{align}
If, on the one hand, \rfr{Domega} is the well known relativistic pericenter advance per orbit, on the other hand, \rfr{DMGE} represents a novel result which amends several incorrect expressions existing in the literature \citep{1977CeMec..15...21R,2005A&A...433..385I, 2007Ap&SS.312..331I}, mainly because based only on \rfr{detadt}.
Indeed, it turns out that \rfr{inte}, integrated over an orbital revolution, does not vanish.
By numerically calculating \rfr{DMGE} with the physical and orbital parameters of some binary, it can be shown that it agrees with the expression obtainable for $\ang{\Delta\mathcal{M}}$ from Equations~(A2.78e)~to~(A2.78f) by \citet[p.~178]{1989racm.book.....S} in which all the three anomalies $f,~E,~\mathcal{M}$ appear.
It should be remarked that \rfr{DMGE} is an exact result in the sense that no a-priori assumptions on $e$ were assumed. It can be shown that, to the zero order in $e$, \rfr{DMGE} is independent of $f_0$.

We will not explicitly display here the analytical expressions for the instantaneous changes $\Delta\kappa^\textrm{GE}\ton{f},\kappa=a,~e,~I,~\Omega,~\omega,~\Delta\mathcal{M}^\textrm{GE}\ton{f}$ because of their cumbersomeness, especially as far as the mean anomaly is concerned. However, $\Delta\kappa^\textrm{GE}\ton{f},\kappa=a,~e,~I,~\Omega,~\omega$ can be found in Equations~(A2.78b)~to~(A2.78d) of \citet[p.~178]{1989racm.book.....S}. Equations~(A2.78e)~to~(A2.78f) of \citet[p.~178]{1989racm.book.....S} allow to obtain the instantaneous shift of the mean anomaly, although in terms of the three anomalies $f,~E,~\mathcal{M}$; instead, our (lengthy) expression contains only the true anomaly $f$. See also Equations~(3.1.102)~to~(3.1.107) of \citet[p.~93]{1991ercm.book.....B}.

\begin{figure*}
\centerline{
\vbox{
\begin{tabular}{c}
\epsfbox{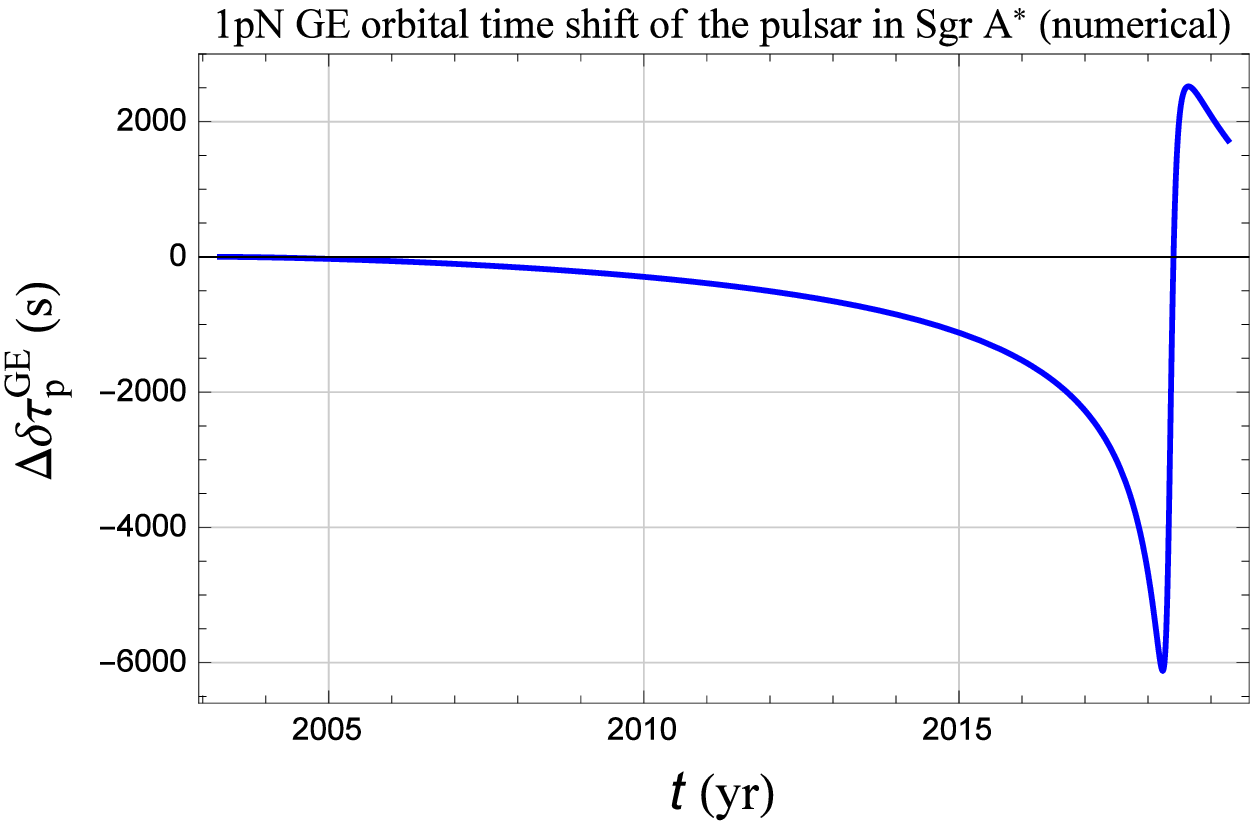}\\
\epsfbox{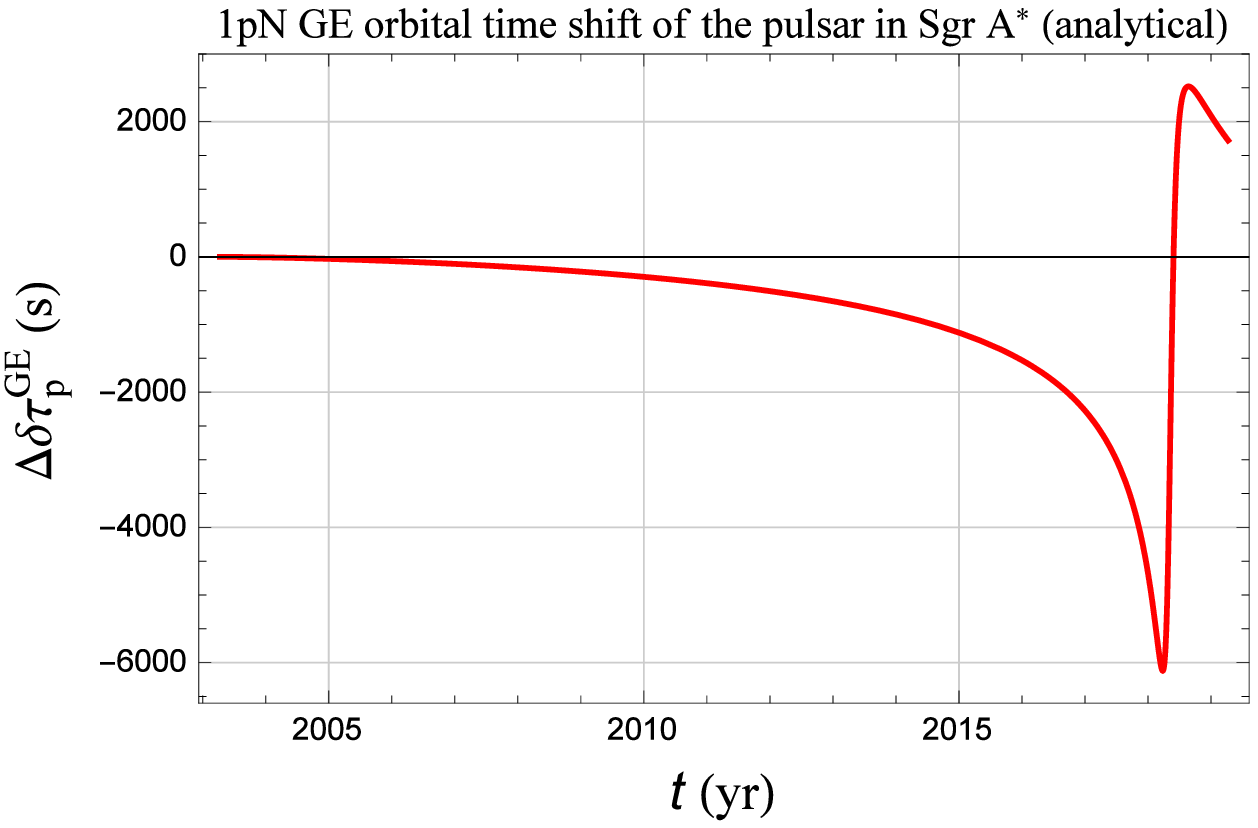}\\
\end{tabular}
}
}
\caption{Upper row, blue curve: $\Delta\delta\tau_\textrm{p}^\textrm{GE}\ton{t}$, in s, of a hypothetical pulsar in Sgr A$^\ast$ as the outcome of the difference between two numerical integrations of the equations of motion in Cartesian coordinates over a time span ranging from, say, $t_0=2003.271$ to $t_0+\Pb$. Both the integrations share the same (Keplerian) initial conditions for $f_0 = 139.72~\textrm{deg}$, corresponding to $t_0=2003.271$, and differ by the 1pN Schwarzschild-like acceleration, which was purposely switched off in one of the two runs. Lower row, red curve: $\Delta\delta\tau_\textrm{p}^\textrm{GE}\ton{t}$, in s,  of a hypothetical pulsar in Sgr A$^\ast$ obtained analytically from \rfrs{Dzf}{fMt} and the instantaneous changes of the Keplerian orbital elements, not displayed in the text, induced by \rfr{AGE}. It turns out that the net shift after a full revolution starting at $t_0=2003.271$ amounts to $\ang{\Delta\delta\tau_\textrm{p}^\textrm{GE}} = 1,722.6948~\textrm{s}$. The orbital configuration of the S2 star, quoted in Table \ref{S2}, was adopted for the putative pulsar in Sgr A$^\ast$.
}\label{Fig1}
\end{figure*}

The net time change per revolution of the pulsar p can be calculated with \rfr{Dzf}
together with \rfrs{DaeIOGE}{DMGE}, by obtaining
\begin{align}
\rp{c^3}{\uppi Gm_\textrm{c}\sI}\ang{\Delta\delta\tau_\textrm{p}^\textrm{GE}}\nonumber \lb{dtGE}& = \rp{6\cos u_0}{\ton{1+e\cos f_0}}+ \\ \nonumber \\
\nonumber &+\rp{\ton{e\co+\cos u_0}}{4 \ton{1-e^2}^{5/2}}\grf{8\ton{-9+2\xi}  +4e^4\ton{-6+7\xi} +e^2\ton{-84+76\xi} + \right.\\ \nonumber\\
\nonumber &+\left.3e\qua{ 8\ton{-7+3\xi}  +e^2\ton{-24+31\xi }   }\cos f_0 + \right.\\ \nonumber \\
&+\left. 3e^2\qua{ 4\ton{-5+4\xi}\cos 2f_0 + e\xi\cos 3 f_0  }}.
\end{align}
It should be noted that \rfr{dtGE} is independent of the semimajor axis $a$, depending only on the shape of the orbit through $e$ and its orientation in space through $I,~\omega$. Furthermore, \rfr{dtGE} does depend on the initial epoch $t_0$ through $f_0$. In the limit $e\rightarrow 0$, \rfr{dtGE} does not vanish, reducing to
\eqi
\ang{\Delta\delta\tau_\textrm{p}^\textrm{GE}} \simeq \rp{4\uppi Gm_\textrm{c}\sin I\ton{-3+\xi}\cos u_0}{c^3} + \mathcal{O}\ton{e}.
\eqf
In view of its cumbersomeness, we will not display here the explicit expression of $\Delta\delta\tau^\textrm{GE}_\textrm{p}\ton{f}$ whose validity was successfully checked by numerically integrating the equations of motion for a fictitious binary system, as shown by Figure \ref{Fig1}; see also Section \ref{psrsgrage}.

We will not deal here with the Shapiro-like propagation delay since it was accurately calculated in the literature; see, e.g., \citet{DamDer86, 1992PhRvD..45.1840D} and references therein.
\subsection{The pulsar in Sgr A$^\ast$ and the gravitoelectric orbital time delay}\lb{psrsgrage}
An interesting, although still observationally unsupported, scenario involves the possibility that radio pulsars orbit the SMBH at the GC in Sgr A$^\ast$; in this case, the unseen companion would be the SMBH itself. Thus, in view of its huge mass, the expected time shift per orbit $\ang{\Delta\delta\tau^\textrm{GE}_\textrm{p}}$ would be quite large.

By considering a hypothetical pulsar with standard mass $m_\textrm{p} = 1.4~\textrm{M}_\odot$ and, say, the same orbital parameters of the main sequence star S2 actually orbiting the Galactic SMBH \citep{2017ApJ...837...30G},
\rfr{dtGE} yields
\eqi
\ang{\Delta\delta\tau^\textrm{GE}_\textrm{p}} = 1,722.6948~\textrm{s}.\lb{megaGE}
\eqf
Figure \ref{Fig1} displays the temporal pattern of $\Delta\delta\tau_\textrm{p}^\textrm{GE}\ton{t}$ for the same hypothetical pulsar calculated both analytically with \rfrs{Dzf}{fMt} applied to \rfr{AGE} and numerically by integrating its equations of motion: their agreement is remarkable. It turns out that
\begin{align}
\left.\Delta\delta\tau_\textrm{p}^\textrm{GE}\right|^\textrm{max} & = 2520.3557~\textrm{s}, \\ \nonumber \\
\left.\Delta\delta\tau_\textrm{p}^\textrm{GE}\right|^\textrm{min} & = -6119.2341~\textrm{s}.
\end{align}

\Rfr{dtGE} allows to find the maximum and minimum values of the net orbital change per revolution of the putative pulsar in Sgr A$^\ast$  by suitably varying  $e,~I,~\omega,~f_0$ within given ranges.
By limiting ourselves to $0 \leq e \leq 0.97$ for convergence reasons of the optimization algorithm adopted, we have
\begin{align}
\ang{\Delta\delta\tau_\textrm{p}^\textrm{GE}}_\textrm{max}  \nonumber & = 1.74521212562\times 10^7~\ton{e_\textrm{max} =0.97,~I_\textrm{max} = 94.98~\textrm{deg},\right.\\ \nonumber\\
&\left.\omega_\textrm{max} = 184.56~\textrm{deg},~f_0^\textrm{max} = 357.23~\textrm{deg}}, \\ \nonumber \\
\ang{\Delta\delta\tau_\textrm{p}^\textrm{GE}}_\textrm{min}  \nonumber & = -1.7568613043\times 10^7~\ton{e_\textrm{min} =0.97,~I_\textrm{min} = 89.95~\textrm{deg},\right.\\ \nonumber \\
&\left.\omega_\textrm{min} = 359.68~\textrm{deg},~f_0^\textrm{min} = 0.28~\textrm{deg}}.
\end{align}

Such huge orbital time delays would be accurately detectable, even by assuming a pessimistic  pulsar timing precision of just $100~\mu\textrm{s}$ \citep{2016ApJ...818..121P,2017IJMPD..2630001G}; more optimistic views point towards precisions of the order of even $1-10~\mu\textrm{s}$ \citep{2016ApJ...818..121P,2017IJMPD..2630001G}.
\section{The 1pN gravitomagnetic Lense-Thirring effect}\lb{LTeffects}
The stationary component of the 1pN field, due to mass-energy currents, is responsible of several aspects of the so-called spin-orbit coupling, or frame-dragging \citep{1986SvPhU..29..215D,1988nznf.conf..573T,2004GReGr..36.2223S,2009SSRv..148...37S}.

The 1pN gravitomagnetic, Lense-Thirring-type, acceleration affecting the relative orbital motion of a generic binary made of two rotating bodies $\textrm{A,~B}$ is \citep{1975PhRvD..12..329B,1989racm.book.....S}
\eqi
{\bds A}_\textrm{LT} = \rp{2G}{c^2 r^3}\qua{ 3\ton{\bds{\mathcal{S}}\bds\cdot\uR}\uR\bds\times\mathbf{v}  + \mathbf{v}\bds\times\bds{\mathcal{S}}  }.\lb{ALT}
\eqf
In general, it is
\eqi \kA\neq\kB,\eqf i.e. the angular momenta of the two bodies are usually not aligned. Furthermore, they are neither aligned with the orbital angular momentum $\bds L$, whose unit vector is given by $\uN$. Finally, also the magnitudes $S^\textrm{A},~S^\textrm{B}$ are, in general, different.

The radial, transverse and out-of-plane components of the gravitomagnetic acceleration, obtained by projecting \rfr{ALT} onto the unit vectors
$\uR,~\uT,~\uN$, turn out to be
\begin{align}
A_{\rho}^\textrm{LT} \lb{ARLT} & = \rp{2G\nk \ton{1+e\csf}^4\bds{\mathcal{S}}\bds\cdot\uN}{c^2 a^2\ton{1-e^2}^{7/2}}, \\ \nonumber \\
A_{\sigma}^\textrm{LT} \lb{ATLT} & = -\rp{2eG\nk \ton{1+e\csf}^3\snf~\bds{\mathcal{S}}\bds\cdot\uN}{c^2 a^2\ton{1-e^2}^{7/2}}, \\ \nonumber \\
A_{\nu}^\textrm{LT} \nonumber \lb{ANLT} & = -\rp{2G\nk\ton{1+e\csf}^3}{c^2 a^2 \ton{1-e^2}^{7/2}}\bds{\mathcal{S}}\bds\cdot\grf{\qua{e\co -\ton{2+3e\csf}\cu}\ul -\right.\\ \nonumber \\
&-\left. \rp{1}{2}\qua{e\so +4\su + 3e\sin\ton{\omega+2f}  }\um  }.
\end{align}

By using \rfrs{ARLT}{ANLT} in \rfr{Dk} and \rfrs{anom}{inte} and integrating them from $f_0$ to $f_0+2\uppi$, it is possible to straightforwardly calculate the 1pN gravitomagnetic net orbital changes for a generic binary arbitrarily oriented in space: they are
\begin{align}
\ang{\Delta a^\textrm{LT}} \lb{DaeMLT}& = \ang{\Delta e^\textrm{LT}}=\ang{\Delta\mathcal{M}^\textrm{LT}} = 0, \\ \nonumber \\
\ang{\Delta I^\textrm{LT}} \lb{DILT}& = \rp{4\uppi G\bds{\mathcal{S}}\bds\cdot\ul}{c^2 \nk a^3\ton{1-e^2}^{3/2}}, \\ \nonumber \\
\ang{\Delta \Omega^\textrm{LT}} \lb{DOLT}& = \rp{4\uppi G\csc I\bds{\mathcal{S}}\bds\cdot\um}{c^2 \nk a^3\ton{1-e^2}^{3/2}}, \\ \nonumber \\
\ang{\Delta \omega^\textrm{LT}} \lb{DoLT}& = -\rp{4\uppi G\bds{\mathcal{S}}\bds\cdot\ton{2\uN + \cot I\um}}{c^2 \nk a^3\ton{1-e^2}^{3/2}}.
\end{align}
It is interesting to remark that, in the case of \rfr{ALT}, both \rfr{detadt} and \rfr{inte} yield vanishing contributions to $\ang{\Delta\mathcal{M}^\textrm{LT}}$.
For previous calculations based on different approaches and formalisms, see, e.g., \citet{1975PhRvD..12..329B, 1988NCimB.101..127D, 2011PhRvD..84l4001I}, and references therein.

\begin{figure*}
\centerline{
\vbox{
\begin{tabular}{c}
\epsfbox{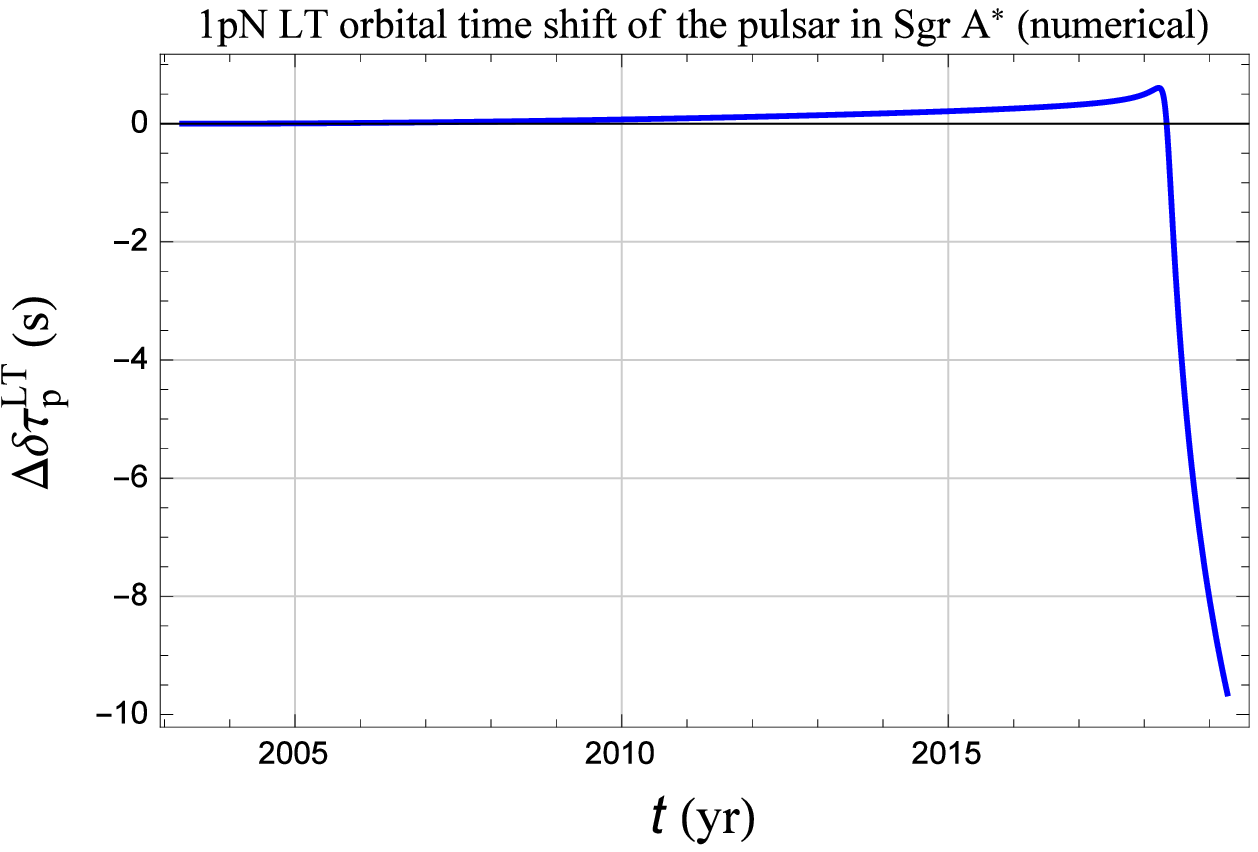}\\
\epsfbox{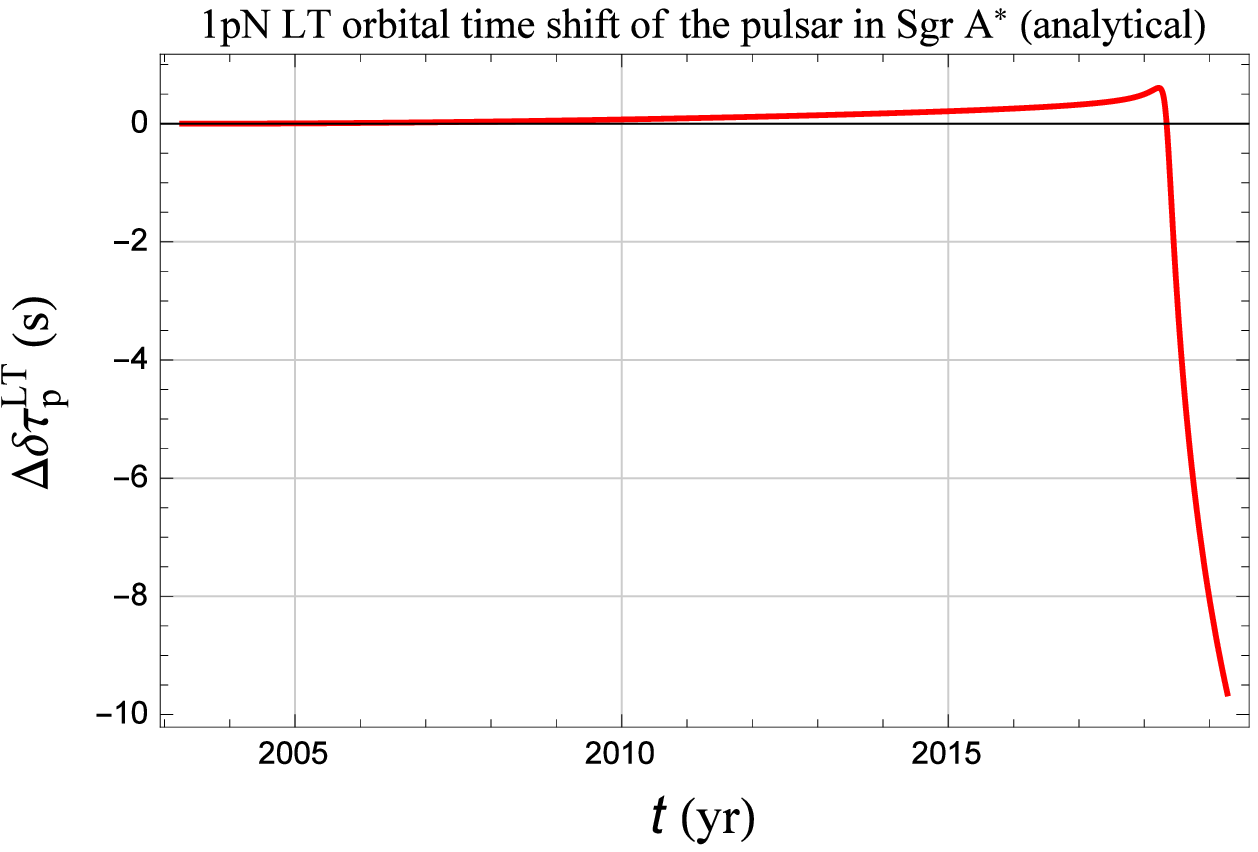}\\
\end{tabular}
}
}
\caption{Upper row, blue curve: $\Delta\delta\tau_\textrm{p}^\textrm{LT}\ton{t}$, in s, of a hypothetical pulsar in Sgr A$^\ast$ as the outcome of the difference between two numerical integrations of the equations of motion in Cartesian coordinates over a time span ranging from, say, $t_0=2003.271$ to $t_0+\Pb$. Both the integrations share the same (Keplerian) initial conditions for $f_0 = 139.72~\textrm{deg}$, corresponding to $t_0=2003.271$, and differ by the 1pN gravitomagnetic acceleration, which was purposely switched off in one of the two runs. Lower row, red curve: $\Delta\delta\tau_\textrm{p}^\textrm{LT}\ton{t}$, in s,  of a hypothetical pulsar in Sgr A$^\ast$ obtained analytically from \rfrs{Dzf}{fMt} and the instantaneous changes of the Keplerian orbital elements, not displayed in the text, induced by \rfr{ALT}. It turns out that the net shift after a full revolution starting at $t_0=2003.271$ amounts to $\ang{\Delta\delta\tau_\textrm{p}^\textrm{LT}} = -9.6439~\textrm{s}$. The orbital configuration of the S2 star, quoted in Table \ref{S2}, was adopted for the putative pulsar in Sgr A$^\ast$ along with $i_\bullet = 20.9~\textrm{deg},~\varepsilon_\bullet = 317.9~\textrm{deg}$.
}\label{Fig2}
\end{figure*}

\Rfr{Dzf}, calculated with \rfrs{DaeMLT}{DoLT}, allows to obtain the net orbit-type time change per revolution of the pulsar p as
\eqi
\ang{\Delta\delta\tau_\textrm{p}^\textrm{LT}} = \rp{4\uppi G m_\textrm{c}}{m_\textrm{tot} c^3 a^2\nk\sqrt{1-e^2}\ton{1+e\cos f_0}}\bds{\mathcal{S}}\bds\cdot\qua{ \ul\cI\sin u_0
-\ton{\um\cI+2\uN\sI}\cos u_0}.\lb{DtauLT}
\eqf
Note that, contrary to \rfr{dtGE}, \rfr{DtauLT} does depend on the semimajor axis as $a^{-1/2}$. As \rfr{dtGE}, also \rfr{DtauLT} depends on $f_0$.
The instantaneous orbital time shift $\Delta\delta\tau_\textrm{p}^\textrm{LT}\ton{f}$ turns out to be too unwieldy to be explicitly displayed here. Its validity was successfully checked by numerically integrating the equations of motion for a fictitious binary system, as shown by Figure \ref{Fig2}; see also Section \ref{psrsgralt}.

The gravitomagnetic propagation time delay is treated in Section \ref{propgLT}.
\subsection{The pulsar in Sgr A$^\ast$ and the Lense-Thirring orbital time delay}\lb{psrsgralt}
Let us, now, consider the so-far hypothetical scenario of an emitting radio pulsar orbiting the SMBH in Sgr A$^\ast$ \citep{2010ApJ...720.1303A,2017IJMPD..2630001G}.

It turns out that, in some relevant astronomical and astrophysical binary systems of interest like the one at hand, the (scaled) angular momentum $\mathcal{S}^\textrm{A/B}$ of one of the bodies is usually much smaller than the other one. Let us assume that the pulsar under consideration has the same characteristics of PSR J0737-3039A.
By assuming \citep{2004ApJ...617L.135M,2005MNRAS.364..635B}
\eqi
I_\textrm{NS}\simeq 10^{38}~\textrm{kg~m}^2\lb{inerzia}
\eqf
for the moment of inertia of a neutron star (NS), the spin of PSR J0737-3039A is
\eqi
S^\textrm{A} \lb{spinA} = 2.8\times 10^{40}~\textrm{kg~m}^2~\textrm{s}^{-1}.
\eqf
%
The angular momentum of a NS of mass $m_\textrm{NS}$ can also be expressed in terms of the dimensionless parameter $\chi_g>0$ as \citep{1999ApJ...512..282L}
\eqi
S^\textrm{NS} = \chi_g\rp{m_\textrm{NS}^2 G}{c}.\lb{SNS}
\eqf
Thus, \rfr{spinA} implies
\eqi
\chi_g^\textrm{A} \lb{miniA} = 0.01755
\eqf
for PSR J0737-3039A.
%
Since for the Galactic SMBH it is \citep{1974JMP....15...46H}
\eqi
S_\bullet = \chi_g \rp{M_{\bullet}^2 G}{c}\simeq 9.68\times 10^{54}~\textrm{kg~m}^2~\textrm{s}^{-1}\lb{sBH}
\eqf
with\footnote{Let us recall that, for a Kerr BH, it must be $\chi_g\leq 1$.} \citep{2016ApJ...818..121P}
\eqi
\chi_g \simeq 0.6,\lb{chig}
\eqf
we have
\eqi
\rp{\mathcal{S}_\textrm{p}}{\mathcal{S}_\bullet} \simeq 6\times 10^{-9}\lb{mini}.
\eqf
Thus, in this case, the dominant contribution to \rfr{ALT} is due to the pulsar's companion c.
As far as the orientation of the SMBH's spin is concerned, we model it as
\eqi
{\bds{\hat{S}}}^\bullet =\sin i_\bullet\cos\varepsilon_\bullet~\uI + \sin i_\bullet\sin\varepsilon_\bullet~\uJ +\cos i_\bullet~\uK.
\eqf
The angles $i_\bullet,~\varepsilon_\bullet$ are still poorly constrained \citep{2009ApJ...697...45B,2011ApJ...735..110B,2016ApJ...827..114Y}, so that we prefer to treat them as free parameters by considering their full ranges of variation
\begin{align}
0\lb{ispin}&\leq i_\bullet \leq 180~\textrm{deg}, \\ \nonumber \\
0\lb{espin}&\leq \varepsilon_\bullet \leq 360~\textrm{deg}.
\end{align}
Also in this case, we assume for our putative pulsar the same orbital parameters of, say, the S2 star.

By using our analytical expression for $\Delta\delta\tau_\textrm{p}^\textrm{LT}\ton{t}$, calculated with $\mathcal{S}^\textrm{p}\rightarrow 0$ in view of \rfr{mini}, one gets
\begin{align}
\left.\Delta\delta\tau^\textrm{LT}_\textrm{p}\right|^\textrm{max} \lb{megaLTmax} &= 0.6054~\textrm{s}~\ton{i_\bullet = 20.9~\textrm{deg},~\varepsilon_\bullet = 317.9~\textrm{deg}},\\ \nonumber\\
\left.\Delta\delta\tau^\textrm{LT}_\textrm{p}\right|^\textrm{min} \lb{megaLTmin}&= -0.6053~\textrm{s}~\ton{i_\bullet= 159.1~\textrm{deg},~\varepsilon_\bullet = 137.9~\textrm{deg}}.
\end{align}
within the assumed ranges of variation for the angles $i_\bullet,~\varepsilon_\bullet$ provided by \rfrs{ispin}{espin}. To this aim, see Figure \ref{Fig2} which shows both the analytical time series, calculated with \rfrs{Dzf}{fMt} applied to \rfr{ALT} for $i_\bullet = 20.9~\textrm{deg},~\varepsilon_\bullet = 317.9~\textrm{deg}$ and the numerically integrated one for the same values of the SMBH's spin axis angles: they are in good agreement. The maximum and the minimum values of the propagation delays for the same orbital configuration of the pulsar are displayed in \rfrs{propLTmax}{propLTmin}.
Let us, now, remove any limitation on the orbital configuration of the pulsar. By restricting ourselves to $0 \leq e \leq 0.97$ for convergence reasons of the optimization algorithm adopted, we have
\begin{align}
\ang{\Delta\delta\tau_\textrm{p}^\textrm{LT}}_\textrm{max}  \nonumber \lb{letimax}& = 411.1823~\textrm{s}~\ton{\Pb^\textrm{max} = 5~\textrm{yr},~e_\textrm{max} =0.97,~I_\textrm{max} = 90~\textrm{deg},\right.\\ \nonumber\\
\nonumber &\left.\omega_\textrm{max} = 180~\textrm{deg},~\Omega^\textrm{max} = 167.21~\textrm{deg},~f_0^\textrm{max} = 180~\textrm{deg},\right.\\ \nonumber \\
&\left. i_\bullet^\textrm{max} = 90~\textrm{deg},~\varepsilon_\bullet^\textrm{max} = 257.21~\textrm{deg}    }, \\ \nonumber \\
\ang{\Delta\delta\tau_\textrm{p}^\textrm{LT}}_\textrm{min}  \nonumber \lb{letimin}& = -293.1933~\textrm{s}~\ton{\Pb^\textrm{min} = 5~\textrm{yr},~e_\textrm{min} =0.97,~I_\textrm{min} = 46.03~\textrm{deg},\right.\\ \nonumber \\
\nonumber &\left.\omega_\textrm{min} = 26.31,~\Omega^\textrm{min} = 14.35~\textrm{deg},~\textrm{deg},~f_0^\textrm{min} = 179.43~\textrm{deg},\right.\\ \nonumber \\
&\left. i_\bullet^\textrm{min} = 159.76~\textrm{deg},~\varepsilon_\bullet^\textrm{min} = 36.41~\textrm{deg}     },\end{align}
where we considered $5~\textrm{yr} \leq \Pb \leq 16~\textrm{yr}$; the ranges of variation assumed for the other parameters $I,~\omega,~\Omega,~f_0,~i_\bullet,~\varepsilon_\bullet$ are the standard full ones.
The values of \rfrs{megaLTmax}{megaLTmin} and \rfrs{letimax}{letimin} should be compared with the expected pulsar timing precision of about $100~\mu\textrm{s}$, or, perhaps, even $1-10~\mu\textrm{s}$ \citep{2016ApJ...818..121P,2017IJMPD..2630001G}.
\subsection{The Lense-Thirring propagation time shift}\lb{propgLT}
The gravitomagnetic propagation delay for a binary with relative separation $r$ and angular momentum $S$ of the primary is \citep{1997JMP....38.2587K, 1999ApJ...514..388W}
\begin{align}
\delta\tau_\textrm{prop}^\textrm{LT} \nonumber \lb{propLT}&= -\rp{2G{\bds S}\bds\cdot\ton{\uK\bds\times\bds r }}{c^4 r (r - \bds{r}\bds\cdot\uK)}=\\ \nonumber \\
&=\rp{2 G S\ton{1+e\csf}\qua{  \cu\ton{ \kx\sO - \ky\cO } + \cI\su\ton{\kx\cO + \ky\sO}  }}{c^4 p\ton{1-\sI\su}}.
\end{align}
According to Figure 2 of \citet{1999ApJ...514..388W}, their unit vector ${\bds K}_0$  agrees with our $\uK$ since it is directed towards the Earth.
About the spin axis of the primary, identified with a BH by \citet{1999ApJ...514..388W}, our $i_\bullet$ coincides with their $\lambda_\bullet$. Instead, their angle $\eta_\bullet$ is reckoned from our unit vector $\ul$, i.e. it is as if \citet{1999ApJ...514..388W} set $\Omega=0$. On the contrary, our angle $\varepsilon_\bullet$ is counted from the reference $x$ direction in the plane of the sky whose unit vector $\uI$, in general, does not coincide with $\ul$. Furthermore, \citet{1999ApJ...514..388W} use the symbol $i$ for the orbital inclination angle, i.e., our $I$.
It is important to notice that, contrary to the orbital time delay of \rfr{DtauLT}, \rfr{propLT} is a short-term effect in the sense that there is no net shift over one orbital revolution.
It is also worth noticing that \rfr{propLT} is of order $\mathcal{O}\ton{c^{-4}}$, while \rfr{DtauLT} is of order  $\mathcal{O}\ton{c^{-3}}$.

As far as the putative scenario of the pulsar in the GC is concerned, the emitting neutron star is considered as the source s of the electromagnetic beam delayed by the angular momentum of the SMBH. Thus, by calculating \rfr{propLT} for a S2-type orbit and with $S = S^\bullet$, it is possible to obtain
\begin{align}
\left.\delta\tau^\textrm{LT}_\textrm{prop}\right|^\textrm{max} \lb{propLTmax}&=0.0195~\textrm{s}~\ton{i_\bullet=90~\textrm{deg},~\varepsilon_\bullet=80.1~\textrm{deg},~f=360~\textrm{deg}},\\ \nonumber\\
\left.\delta\tau^\textrm{LT}_\textrm{prop}\right|^\textrm{min} \lb{propLTmin}&= -0.0213~\textrm{s}~\ton{i_\bullet=90~\textrm{deg},~\varepsilon_\bullet=235.5~\textrm{deg},~f=18.5~\textrm{deg}}.
\end{align}
Such values are one order of magnitude smaller than \rfrs{megaLTmax}{megaLTmin} for the orbital time delay calculated with the same orbital configuration of the pulsar.
It turns out that values similar to those of \rfrs{propLTmax}{propLTmin} are obtained by discarding the S2 orbital configuration for the pulsar:
\begin{align}
\left.\delta\tau_\textrm{prop}^\textrm{LT}\right|^\textrm{max}  \nonumber \lb{maxleti}& =
0.0105~\textrm{s}~
\ton{
\Pb^\textrm{max} = 10.5~\textrm{yr},
~e_\textrm{max} =0.97,
~I_\textrm{max} = 54.98~\textrm{deg},\right.\\ \nonumber\\
\nonumber &\left.
\omega_\textrm{max} = 70.09~\textrm{deg},
~\Omega^\textrm{max} = 70.33~\textrm{deg},
~f^\textrm{max} = 69.94~\textrm{deg},\right.\\ \nonumber \\
&\left.
i_\bullet^\textrm{max} = 55.11~\textrm{deg},
~\varepsilon_\bullet^\textrm{max} = 72.19~\textrm{deg}
}, \\ \nonumber \\
\left.\delta\tau_\textrm{prop}^\textrm{LT}\right|^\textrm{min}  \nonumber \lb{minleti}& =
-0.0375~\textrm{s}~\ton{
\Pb^\textrm{min} = 5~\textrm{yr},
~e_\textrm{min} =0.97,
~I_\textrm{min} = 0~\textrm{deg},\right.\\ \nonumber \\
\nonumber &\left.
\omega_\textrm{min} = 325.24~\textrm{deg},
~\Omega^\textrm{min} = 80.08~\textrm{deg},
~f^\textrm{min} = 0~\textrm{deg},\right.\\ \nonumber \\
&\left. i_\bullet^\textrm{min} = 90.01~\textrm{deg},
~\varepsilon_\bullet^\textrm{min} = 135.3~\textrm{deg}
}.\end{align}
\section{The quadrupole-induced effect}\lb{Qeffects}
If both the bodies of an arbitrary binary system are axisymmetric about their spin axes $\bds{\hat{S}}^\textrm{A/B}$, a further non-central relative acceleration arises; it is \citep{1975PhRvD..12..329B}
\eqi
\rp{2r^4}{3\mu }{\bds A}_{J_2} = J_2^\textrm{A}R_\textrm{A}^2\grf{\qua{5\ton{\Kr}^2 - 1}\uR - 2\ton{\Kr}\kA} +{\textrm{A}\leftrightarrows\textrm{B}},\lb{AJ2}
\eqf
in which the first even zonal parameter $J_2^\textrm{A/B}$ is dimensionless.
In the notation of \citet{1975PhRvD..12..329B}, their $J_2^\textrm{A/B}$ parameter is not dimensionless as ours, being dimensionally an area because it corresponds to our $J_2^\textrm{A/B} R_\textrm{A/B}^2$. Furthermore, \citet{1975PhRvD..12..329B} introduce an associated dimensional quadrupolar parameter $\Delta I^\textrm{A/B}$, having the dimensions of a moment of inertia, which is connected to our $J_2^\textrm{A/B}$ by
\eqi
J_2^\textrm{A/B} = \rp{\Delta I^\textrm{A/B}}{M_\textrm{A/B}~R^2_\textrm{A/B}}.
\eqf
Thus, $\Delta I^\textrm{A/B}$ corresponds to the dimensional quadrupolar parameter $Q_2^\textrm{A/B}$ customarily adopted when astrophysical compact objects like neutron stars and black holes are considered \citep{1999ApJ...512..282L,
2014PhRvD..89d4043W}, up to a minus sign, i.e.
\eqi
J_2^\textrm{A/B}=-\rp{Q^\textrm{A/B}_2}{M_\textrm{A/B}~R_\textrm{A/B}^2}.\lb{J2Q}
\eqf
Thus,
\rfr{AJ2} can be written as
\eqi
\rp{2r^4}{3G}{\bds A}_{Q_2} = {\mathcal{Q}}_2^\textrm{A}\grf{\qua{1 - 5\ton{\Kr}^2 }\uR + 2\ton{\Kr}\kA }+{\textrm{A}\leftrightarrows\textrm{B}}\lb{AQ2}.
\eqf

Projecting \rfr{AJ2} onto the radial, tranvserse  and out-of-plane  unit vectors  $\uR,~\uT,~\uN$ provides us with
\begin{align}
\rp{2a^4\ton{1-e^2}^4}{3\mu\ton{1+e\csf}^4}A_{\rho}^{J_2} \lb{ARJ2}& = J_2^\textrm{A}R_\textrm{A}^2\grf{3\qua{\cu\ton{\Kl} + \su\ton{\Km}}^2 - 1}+{\textrm{A}\leftrightarrows\textrm{B}}, \\ \nonumber \\
-\rp{a^4\ton{1-e^2}^4}{3\mu\ton{1+e\csf}^4}A_{\sigma}^{J_2} \lb{ATJ2}\nonumber & = J_2^\textrm{A}R_\textrm{A}^2\qua{\cu\ton{\Kl} + \su\ton{\Km}}\qua{\cu\ton{\Km} - \su\ton{\Kl}}+\\ \nonumber \\
&+ {\textrm{A}\leftrightarrows\textrm{B}}, \\ \nonumber \\
-\rp{a^4\ton{1-e^2}^4}{3\mu\ton{1+e\csf}^4}A_{\nu}^{J_2} \lb{ANJ2}& = J_2^\textrm{A}R_\textrm{A}^2\qua{\cu\ton{\Kl} + \su\ton{\Km}}\ton{\Kn}+{\textrm{A}\leftrightarrows\textrm{B}}.
\end{align}

A straightforward consequence of \rfrs{ARJ2}{ANJ2} is the calculation of the net quadrupole-induced shifts per revolution of the Keplerian orbital elements by means of \rfr{Dk} and \rfrs{anom}{inte}, which turn out to be
\begin{align}
\ang{\Delta a^{J_2}}&=\ang{\Delta e^{J_2}}\lb{DaeJ2}=0, \\ \nonumber \\
-\rp{p^2}{3\uppi}\ang{\Delta I^{J_2}} \lb{DIJ2}& = J_2^\textrm{A}R^2_\textrm{A}\ton{\Kl}\ton{\Kn}+{\textrm{A}\leftrightarrows\textrm{B}}, \\ \nonumber \\
-\rp{p^2}{3\uppi}\ang{\Delta\Omega^{J_2}} \lb{DOJ2}& = J_2^\textrm{A}R^2_\textrm{A}\ton{\Km}\ton{\Kn}\csc I+{\textrm{A}\leftrightarrows\textrm{B}}, \\ \nonumber \\
\nonumber \rp{2p^2}{3\uppi}\ang{\Delta\omega^{J_2}} \lb{DoJ2}& = J_2^\textrm{A}R^2_\textrm{A}\grf{2 - 3\qua{\ton{\Kl}^2+\ton{\Km}^2}+2\ton{\Km}\ton{\Kn}\cot I}+ \\ \nonumber \\
&+ {\textrm{A}\leftrightarrows\textrm{B}}, \\ \nonumber \\
\rp{2a^2\ton{1-e^2}^3}{3\uppi\ton{1+e\cos f_0}^3}\ang{\Delta\mathcal{M}^{J_2}} \nonumber \lb{DMJ2}& = J_2^\textrm{A}R^2_\textrm{A}\grf{2 - 3\qua{\ton{\Kl}^2+\ton{\Km}^2} - \right.\\ \nonumber \\
\nonumber &-\left. 3\qua{ \ton{\Kl}^2-\ton{\Km}^2 }\cos 2u_0 - 6\ton{\Kl}\ton{\Km}\sin 2u_0 } +\\ \nonumber \\
& +{\textrm{A}\leftrightarrows\textrm{B}}.
\end{align}
Also \rfr{DMJ2}, as \rfr{DMGE} for the Schwarzschild-like 1pN acceleration, is a novel result which amends the incorrect formulas  widely disseminated in the literature \citep{2004Tapleyetal,2005ormo.book.....R,2005som..book.....C,2008orbi.book.....X,2011PhRvD..84l4001I}; indeed, it turns out that, in the case of \rfr{AJ2},  \rfr{inte} does not vanish when integrated over a full orbital revolution. Furthermore, contrary to almost all of the other derivations existing in the literature, \rfr{DMJ2} is quite general since it holds for a two-body system with generic quadrupole mass moments arbitrarily oriented in space, and characterized by a general orbital configuration. The same remark holds also for \rfrs{DaeJ2}{DoJ2}; cfr. with the corresponding (correct) results by \citet{2011PhRvD..84l4001I} in the case of a test particle orbiting an oblate primary.

\begin{figure*}
\centerline{
\vbox{
\begin{tabular}{c}
\epsfbox{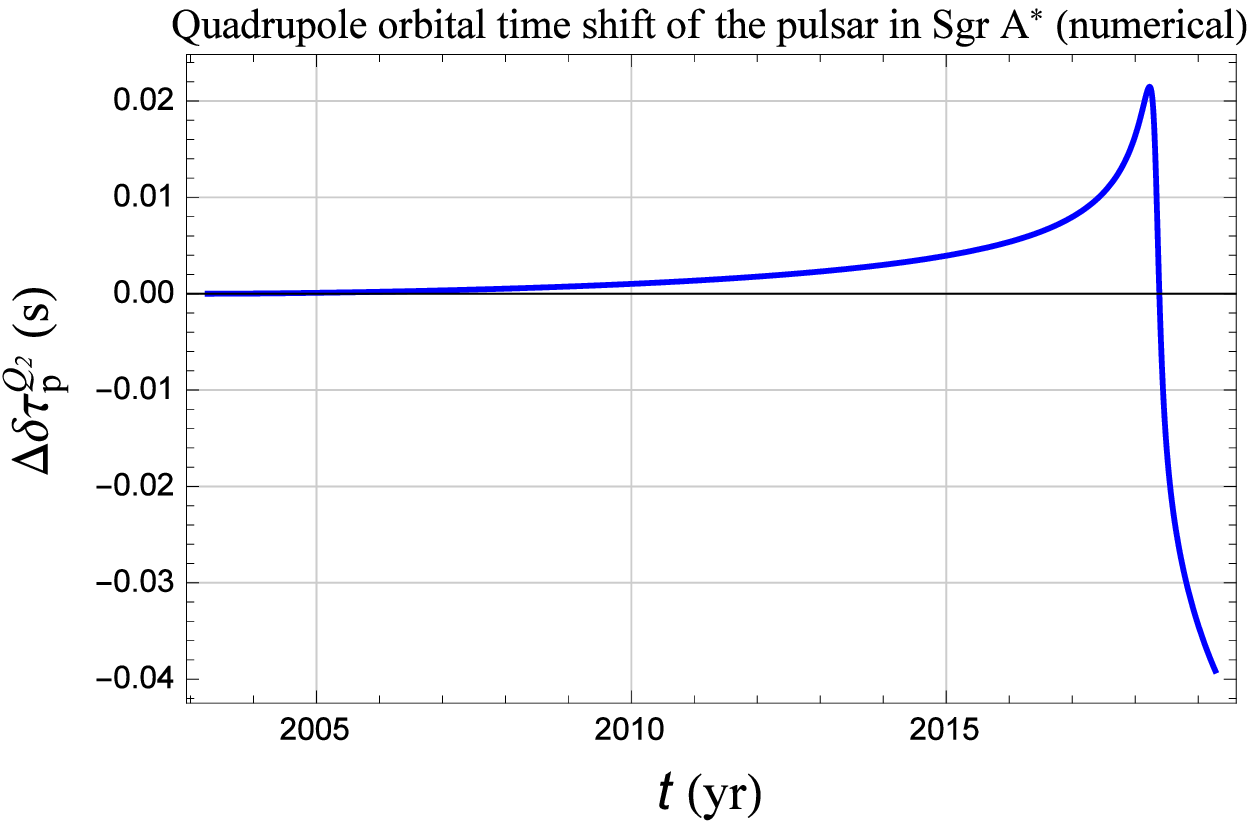}\\
\epsfbox{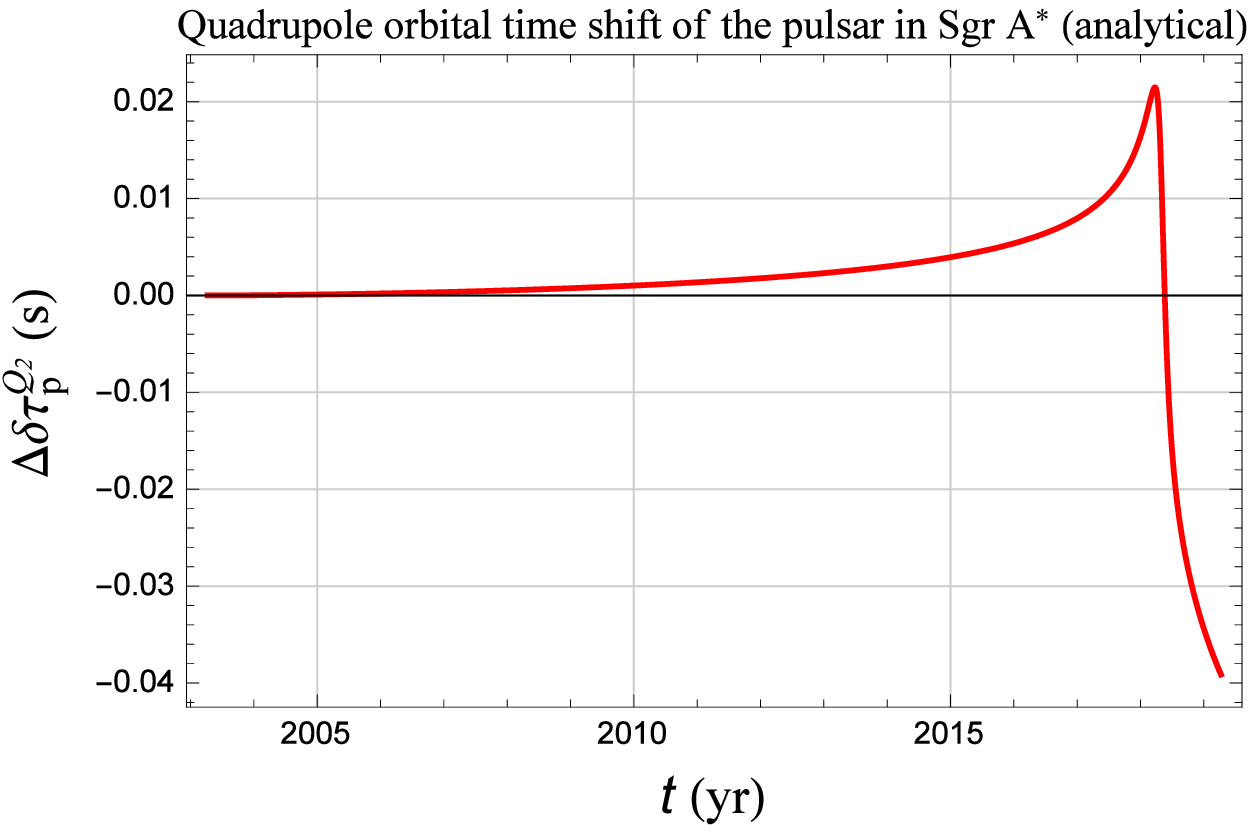}\\
\end{tabular}
}
}
\caption{Upper row, blue curve: $\Delta\delta\tau_\textrm{p}^{Q_2}\ton{t}$, in s, of a hypothetical pulsar in Sgr A$^\ast$ as the outcome of the difference between two numerical integrations of the equations of motion in Cartesian coordinates over a time span ranging from, say, $t_0=2003.271$ to $t_0+\Pb$. Both the integrations share the same (Keplerian) initial conditions for $f_0 = 139.72~\textrm{deg}$, corresponding to $t_0=2003.271$, and differ by the quadrupole-induced acceleration, which was purposely switched off in one of the two runs. Lower row, red curve: $\Delta\delta\tau_\textrm{p}^{Q_2}\ton{t}$, in s,  of a hypothetical pulsar in Sgr A$^\ast$ obtained analytically from \rfrs{Dzf}{fMt} and the instantaneous changes of the Keplerian orbital elements, not displayed in the text, induced by \rfr{AJ2}. It turns out that the net shift after a full revolution starting at $t_0=2003.271$ amounts to $\ang{\Delta\delta\tau_\textrm{p}^{Q_2}} = -0.0391~\textrm{s}$. The orbital configuration of the S2 star, quoted in Table \ref{S2}, was adopted for the putative pulsar in Sgr A$^\ast$ along with $i_\bullet = 146.7~\textrm{deg},~\varepsilon_\bullet = 148.8~\textrm{deg}$.
}\label{Fig3}
\end{figure*}

According to \rfr{Dzf} and \rfrs{DaeJ2}{DMJ2}, the net orbit-like time change of the pulsar p after one orbital revolution is
\begin{align}
\rp{2m_\textrm{tot}ca\ton{1-e^2}\ton{1+e\cos f_0}}{3\uppi m_\textrm{c} }\ang{\Delta\delta\tau_\textrm{p}^{J_2}} \lb{DtauJ2}\nonumber &=   J_2^\textrm{p}R^2_\textrm{p}\qua{2 - 3\ton{\Klp}^2 - 3\ton{\Kmp}^2  }\sI\cos u_0  + \\ \nonumber \\
\nonumber & +  2J_2^\textrm{p}R^2_\textrm{p}\cI\ton{\Knp}\qua{ \ton{\Kmp}\cos u_0-\ton{\Klp}\sin u_0  } -\\ \nonumber \\
\nonumber &-\rp{J_2^\textrm{p}R^2_\textrm{p}\ton{1+e\cos f_0}^4}{\ton{1-e^2}^{5/2}}\sI\ton{e\co + \cos u_0 }\cdot\\ \nonumber \\
\nonumber &\cdot \grf{ -2 + 3\qua{\ton{\Klp}^2+\ton{\Kmp}^2}  + \right.\\ \nonumber \\
\nonumber &+\left.3\qua{\ton{\Klp}^2 - \ton{\Kmp}^2 }\cos 2u_0 + \right. \\ \nonumber \\
& + \left. 6\ton{\Klp}\ton{\Kmp}\sin 2u_0   }  +{\textrm{p}\leftrightarrows\textrm{c}}.
\end{align}
It turns out that \rfr{DtauJ2} does not vanish in the limit $e\rightarrow 0$. If, on the one hand, \rfr{DtauJ2} depends of $f_0$ as \rfr{dtGE} and \rfr{DtauLT}, on the other hand, it depends on the orbital semimajor axis through $a^{-1}$.
As far as $\Delta\delta\tau^{J_2}_\textrm{p}\ton{f}$ is concerned, it will not be displayed explicitly because it is far too ponderous. Also in this case, a numerical integration of the equations of motion for a fictitious binary system, displayed in Figure \ref{Fig3}, confirmed our analytical result for the temporal pattern of $\Delta\delta\tau^{J_2}_\textrm{p}\ton{f}$; see also Section \ref{orbQBH}.

The propagation time delay is dealt with in Section \ref{propQ}.
\subsection{The pulsar in Sgr A$^\ast$ and the quadrupole-induced orbital time delay}\lb{orbQBH}
A rotating NS acquires a non-zero quadrupole moment given by \citep{1999ApJ...512..282L}
\eqi Q_2^\textrm{NS} = q\rp{m_\textrm{NS}^3 G^2}{c^4};\lb{QNS}\eqf
the absolute values of the dimensionless parameter $q<0$ ranges from $0.074$ to $3.507$ for a variety of Equations of State (EOSs) and $m_\textrm{NS} = 1.4~\textrm{M}_\odot$; cfr. Table 4 of \citet{1999ApJ...512..282L}.
It is interesting to note that \citet{1999ApJ...512..282L} find the relation
\eqi
q \simeq -\alpha\chi_g^2,\lb{chiq}
\eqf
where the parameter $\alpha$ of the fit performed by \citet{1999ApJ...512..282L} depends on both the mass of the neutron star and the EOS used. According to Table 7 of \citet{1999ApJ...512..282L},  it is
\begin{align}
\alpha^\textrm{max} \lb{amax1.4}& = 7.4~\ton{m_\textrm{NS}=1.4~\textrm{M}_\odot,~\textrm{EOS~L}}, \\ \nonumber \\
\alpha^\textrm{min} \lb{amin1.4}& = 2.0~\ton{m_\textrm{NS}=1.4~\textrm{M}_\odot,~\textrm{EOS~G}}
\end{align}
for some of the EOSs adopted by \citet{1999ApJ...512..282L}.
In the case of PSR J0737-3039A, \rfr{QNS} yields
\eqi
Q_2^\textrm{A} \lb{QA} = {q}_\textrm{A}~1.04\times 10^{37}~\textrm{kg~m}^2.
\eqf
According to \rfr{miniA} and \rfr{chiq}, it is
\eqi
q_\textrm{A} \lb{qA} = -\alpha_\textrm{A}~3.1\times 10^{-4}.
%
\eqf
As a consequence of the \virg{no-hair} or uniqueness theorems \citep{1972CMaPh..25..152H,1967PhRv..164.1776I,1975PhRvL..34..905R}, the quadrupole moment of a BH is uniquely determined  by its mass and spin according to \citep{1970JMP....11.2580G,1974JMP....15...46H}
\eqi
Q_2^\bullet = -\rp{S^2_\bullet}{c^2 M_\bullet};\lb{qBH}
\eqf
in the case of the SMBH in Sgr A$^\ast$, it is ($\chi_g=0.6$)
\eqi
Q_2^\bullet = -1.2\times 10^{56}~\textrm{kg~m}^2\lb{QBH}.
\eqf
\Rfr{QA} and \rfr{QBH} imply that, in the GC,
\begin{align}
Q_2^\textrm{p}\rp{m_\textrm{c}}{m_\textrm{p}} \lb{miniq1}& = q_\textrm{p}~3.8\times 10^{43}~\textrm{kg~m}^2, \\ \nonumber \\
Q_2^\bullet \lb{miniq2}& = -1.2\times 10^{56}~\textrm{kg~m}^2,
\end{align}
so that the quadrupole of a hypothetical emitting neutron star p orbiting the SMBH in Sgr A$^\ast$ can be completely neglected with respect to the quadrupole of the latter one in any practical calculation.

According to our analytical expression for $\Delta\delta\tau^{Q_2}_\textrm{p}\ton{t}$ applied to a pulsar moving along a S2-type orbit, in view of the ranges of variation assumed in \rfrs{ispin}{espin} for $i_\bullet,~\varepsilon_\bullet$, it is
\begin{align}
\left.\Delta\delta\tau^{Q_2}_\textrm{p}\right|^\textrm{max} \lb{megaQ2max} &= 0.0215~\textrm{s}~\ton{i_\bullet = 146.7~\textrm{deg},~\varepsilon_\bullet = 148.8~\textrm{deg}},\\ \nonumber \\
\left.\Delta\delta\tau^{Q_2}_\textrm{p}\right|^\textrm{min} \lb{megaQ2min} &= -0.0393~\textrm{s}~\ton{i_\bullet = 30.5~\textrm{deg},~\varepsilon_\bullet = 331.6~\textrm{deg}}.
\end{align}
See Figure \ref{Fig3} which displays the outcome of a numerical integration of the equations of motion of the pulsar considered for $i_\bullet = 146.7~\textrm{deg},~\varepsilon_\bullet = 148.8~\textrm{deg}$, and the corresponding analytical time series calculated by means of \rfrs{Dzf}{fMt} applied to \rfr{AJ2}: they agree quite well. The maximum and minimum values of the propagation delay for the same orbital configuration of the pulsar are in \rfrs{maxQBH}{minQBH}. By removing the restrictions on the orbit of the pulsar and assuming the same ranges of variation for $\Pb,~e,~I,~\Omega,~\omega,~f_0,~i_\bullet,~\varepsilon_\bullet$ as in Section \ref{psrsgralt}, apart from $0\leq e\leq 0.96$ for convergence issues of the optimization algorithm adopted, it is possible to obtain
\begin{align}
\ang{\Delta\delta\tau_\textrm{p}^{Q_2}}_\textrm{max}  \nonumber \lb{qu2max}& =
1392.3665~\textrm{s}~
\ton{
\Pb^\textrm{max} = 5~\textrm{yr},
~e_\textrm{max} =0.96,
~I_\textrm{max} = 90~\textrm{deg},\right.\\ \nonumber\\
\nonumber &\left.
\omega_\textrm{max} = 180~\textrm{deg},
~\Omega^\textrm{max} = 17.89~\textrm{deg},
~f_0^\textrm{max} = 0~\textrm{deg},\right.\\ \nonumber \\
&\left.
i_\bullet^\textrm{max} = 90~\textrm{deg},
~\varepsilon_\bullet^\textrm{max} = 17.89~\textrm{deg}
}, \\ \nonumber \\
\ang{\Delta\delta\tau_\textrm{p}^{Q_2}}_\textrm{min}  \nonumber \lb{qu2min}& =
-696.1481~\textrm{s}~\ton{
\Pb^\textrm{min} = 5~\textrm{yr},
~e_\textrm{min} =0.96,
~I_\textrm{min} = 90.05~\textrm{deg},\right.\\ \nonumber \\
\nonumber &\left.
\omega_\textrm{min} = 180.18~\textrm{deg},
~\Omega^\textrm{min} = 37.28~\textrm{deg},
~f_0^\textrm{min} = 0~\textrm{deg},\right.\\ \nonumber \\
&\left. i_\bullet^\textrm{min} = 179.75~\textrm{deg},
~\varepsilon_\bullet^\textrm{min} = 0.89~\textrm{deg}
}.\end{align}
The bounds of \rfrs{megaQ2max}{megaQ2min} and \rfrs{qu2max}{qu2min} can be compared with the minimum and maximum values of the gravitomagnetic orbital shift of \rfrs{megaLTmin}{megaLTmax} for an S2-type orbit, which are about one order of magnitude larger than \rfrs{megaQ2max}{megaQ2min}, and \rfrs{letimax}{letimin}, which, instead, are smaller than \rfrs{qu2max}{qu2min}. Furthermore, the values of \rfrs{megaQ2max}{megaQ2min} and \rfrs{qu2max}{qu2min} seem to be potentially measurable in view of the expected pulsar timing precision of about $100~\mu\textrm{s}$, or, perhaps, even $1-10~\mu\textrm{s}$ \citep{2016ApJ...818..121P,2017IJMPD..2630001G}.
\subsection{The quadrupole-induced propagation time shift}
\begin{figure*}
\centerline{
\vbox{
\begin{tabular}{c}
\epsfysize= 10.0 cm\epsfbox{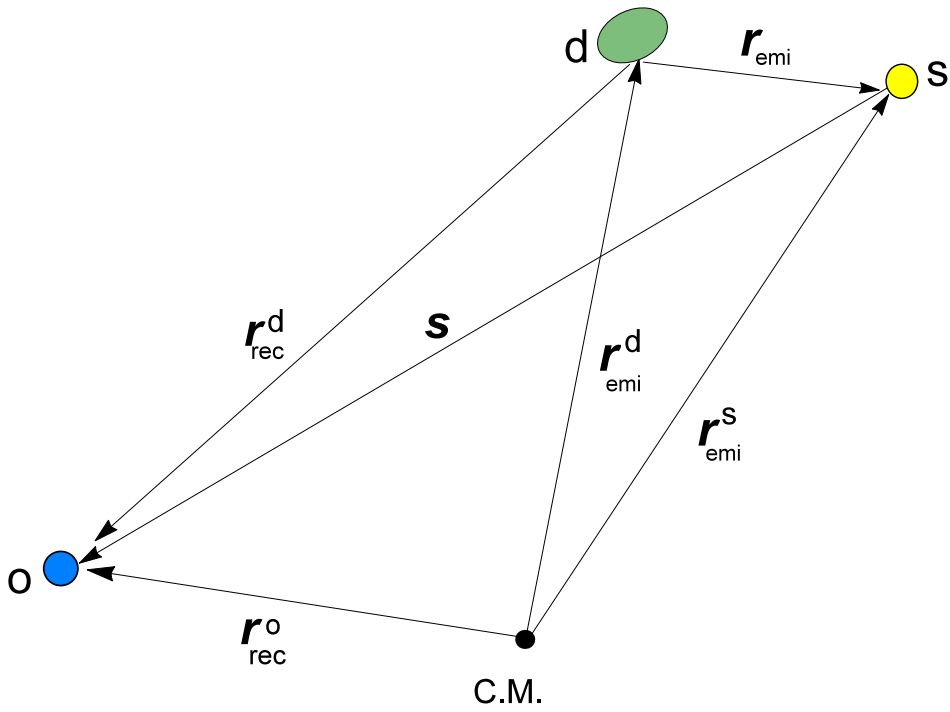}\\
\end{tabular}
}
}
\caption{Vectors connecting the source s and the deflector d at $t_\textrm{emi}$, and the observer o at $t_\textrm{rec}$. The symbols adopted differ from those used in, e.g., \citet{2011CQGra..28a5009Z}. The origin $O$ coincides with the binary's center of mass (C.~M.).
}\label{dise}
\end{figure*}

\lb{propQ}
The propagation delay $\delta\tau_\textrm{prop}^{J_2}$ due to the quadrupole mass moment is rather complicated to be analytically calculated; see, e.g., \citet{1991SvA....35..523K,1997JMP....38.2587K,2003AJ....125.1580K,2011CQGra..28a5009Z}. No explicit expressions analogous to the simple one of \rfr{propLT} for frame-dragging exist in the literature. Here, we will obtain an analytical formula for $\delta\tau_\textrm{prop}^{J_2}$ which will be applied to the double pulsar and the pulsar-Sgr A$^\ast$ systems. The approach by \citet{2011CQGra..28a5009Z} will be adopted by adapting it to the present scenario.
In the following, the subscripts $\textrm{d,~s,~o}$ will denote the deflector, the source, and the observer, respectively.
In the case of, say, the double pulsar, d is the pulsar B, while s is the currently visible pulsar A; in the pulsar-Sgr A$^\ast$ scenario, d is the SMBH and s is the hypothetical pulsar p orbiting it.
See Figure \ref{dise} for the following vectors connecting  $\textrm{d,~s,~o}$. The origin $O$ is at the binary's center of mass, so that
\eqi
{\bds r}^\textrm{s}_\textrm{emi} \doteq {\bds r}^\textrm{s}(t_\textrm{emi})
\eqf
is the barycentric position vector of the source s at the time of emission $t_\textrm{emi}$,
\eqi
{\bds r}^\textrm{d}_\textrm{emi} \doteq {\bds r}^\textrm{d}(t_\textrm{emi})
\eqf
is the barycentric position vector of the deflector d at  $t_\textrm{emi}$,
\eqi
{\bds r}_\textrm{emi} = {\bds r}^\textrm{s}_\textrm{emi}-{\bds r}^\textrm{d}_\textrm{emi}
\eqf
is the relative position vector of the source s with respect to the deflector d at $t_\textrm{emi}$. Thus, to the Newtonian order, it is
\begin{align}
{\bds r}^\textrm{d}_\textrm{emi} & \simeq -\rp{m_\textrm{s}}{m_\textrm{tot}}~{\bds r}_\textrm{emi}, \\ \nonumber \\
{\bds r}^\textrm{s}_\textrm{emi} & \simeq \rp{m_\textrm{d}}{m_\textrm{tot}}~{\bds r}_\textrm{emi},
\end{align}
where $m_\textrm{s},~m_\textrm{d}$ are the masses of source and deflector, respectively.
Furthermore,
\eqi
{\bds r}_\textrm{rec}^\textrm{o}\doteq {\bds r}^\textrm{o}(t_\textrm{rec})
\eqf
is the barycentric position vector of the observer o at the time of reception $t_\textrm{rec}$,
\eqi
{\bds r}_\textrm{rec}^\textrm{d}={\bds r}_\textrm{rec}^\textrm{o}-{\bds r}_\textrm{emi}^\textrm{d}
\eqf
is the position vector of the observer o at $t_\textrm{rec}$ with respect to the deflector d at $t_\textrm{emi}$, and
\eqi
\bds s={\bds r}_\textrm{rec}^\textrm{o}-{\bds r}_\textrm{emi}^\textrm{s}={\bds r}_\textrm{rec}^\textrm{d}-{\bds r}_\textrm{emi}
\eqf
is the position vector of the observer o at $t_\textrm{rec}$ with respect to the source s at $t_\textrm{emi}$.
With our conventions for the coordinate axes, it is
\eqi
{\bds r}_\textrm{rec}^\textrm{o} = D~\uK,
\eqf
where $D$ is the distance of the binary at $t_\textrm{emi}$ from us at $t_\textrm{rec}$, which is usually much larger than the size $r_\textrm{emi}$ of the binary's orbit. Thus, the following simplifications  can be safely made
\begin{align}
\bds s &= {\bds r}_\textrm{rec}^\textrm{o}-{\bds r}_\textrm{emi}^\textrm{s} = D~\uK - \rp{m_\textrm{d}}{m_\textrm{tot}}~{\bds r}_\textrm{emi} \simeq D~\uK, \\ \nonumber \\
\us \lb{kappas}&\simeq \uK, \\ \nonumber \\
{\bds r}_\textrm{rec}^\textrm{d} & = {\bds r}_\textrm{rec}^\textrm{o}-{\bds r}_\textrm{emi}^\textrm{d} =D~\uK + \rp{m_\textrm{s}}{m_\textrm{tot}}~{\bds r}_\textrm{emi} \simeq D~\uK.
\end{align}
To order, $\mathcal{O}\ton{c^{-2}}$, the impact parameter vector can be calculated as \citep{2011CQGra..28a5009Z}
\eqi
{\bds \ell}^\textrm{d} \simeq \us\bds\times\ton{{\bds r}_\textrm{emi}\bds\times\us} = {\bds{r}}_\textrm{emi} -\us\ton{{\bds r}_\textrm{emi}\bds\cdot\us}.
\eqf
In view of \rfr{kappas}, it turns out that ${\bds \ell}^\textrm{d}$, evaluated onto the unperturbed Keplerian ellipse, lies in the plane of the sky, being made of the $x,~y$ components of $\uR$ scaled by \rfr{Kepless}.

The coefficients of Equations~(A.18)~to~(A.20) of \citet{2011CQGra..28a5009Z}
\begin{align}
\mathcal{E}_\textrm{d} & = \rp{\us \bds\cdot {\bds r}_\textrm{emi} }{r_\textrm{emi}^3} - \rp{\us\bds\cdot {\bds r}_\textrm{rec}^\textrm{d} }{{r_\textrm{rec}^\textrm{d}}^3} , \\ \nonumber \\
\mathcal{F}_\textrm{d} & = \ell^\textrm{d}\ton{\rp{1 }{r_\textrm{emi}^3} - \rp{1}{{r_\textrm{rec}^\textrm{d}}^3}}, \\ \nonumber \\
\mathcal{V}_\textrm{d} & = -\rp{1}{{\ell^\textrm{d}}^2}\ton{ \rp{\us \bds\cdot {\bds r}_\textrm{emi} }{r_\textrm{emi}} - \rp{\us \bds\cdot {\bds r}_\textrm{rec}^\textrm{d} }{r_\textrm{rec}^\textrm{d}} },
\end{align}
required to calculate $\delta\tau_\textrm{prop}^{J_2}$,
can be approximated to
\begin{align}
\mathcal{E}_\textrm{d} \lb{Ed} & \simeq \rp{\us \bds\cdot {\bds r}_\textrm{emi} }{r_\textrm{emi}^3}, \\ \nonumber \\
\mathcal{F}_\textrm{d} \lb{Fd} & \simeq \rp{\ell^\textrm{d}}{r_\textrm{emi}^3} , \\ \nonumber \\
\mathcal{V}_\textrm{d} \lb{Vd} & \simeq -\rp{1}{{\ell^\textrm{d}}^2}\ton{ \rp{\us \bds\cdot {\bds r}_\textrm{emi} }{r_\textrm{emi}} - 1 }.
\end{align}

The rotation matrix which brings the deflector's symmetry axis from $\uK$ to a generic position in space characterized by the usual polar angles $i,~\varepsilon$ is
\eqi
\qua{{\textsf{R}}}_{ij} =
\left(
\begin{array}{ccc}
\ce \cis  & -\se  & \ce \sis  \\
\se \cis & \ce  &  \se \sis \\
-\sis & 0 & \cis \\
\end{array}
\right)
\eqf
It is made of an anticlockwise rotation by an amount $i$ around $\uJ$, followed by an anticlockwise rotation by an amount $\varepsilon$ around $\uK$.
The symmetric trace-free quadrupole moment  of the deflector \citep{2011CQGra..28a5009Z}
\eqi
{\textsf{M}}^\textrm{d} =
\rp{1}{3}m_\textrm{d} R_\textrm{d}^2 J_2^\textrm{d}~{\textsf{R}}~{\textsf{D}}[1,1,-2]~{\textsf{R}}^\mathrm{T},
\eqf
where $\textsf{D}\qua{\ast,\ast,\ast}$ denotes the diagonal matrix along with the associated entries,
becomes
\eqi
\qua{{\textsf{M}}^\textrm{d}}_{ij} = m_\textrm{d} R_\textrm{d}^2 J_2^\textrm{d}~
\left(
\begin{array}{ccc}
\rp{1}{3} -\cos^2\varepsilon\sin^2 i  & -\rp{1}{2}\sin^2 i\see  & -\rp{1}{2}\ce\siss  \\
-\rp{1}{2}\sin^2 i\see &  -\rp{2}{3} +\cos^2 \varepsilon +\cos^2 i\sin^2\varepsilon   &  -\rp{1}{2}\siss\se \\
 -\rp{1}{2}\ce\siss &  -\rp{1}{2}\siss\se & \rp{1}{3}-\cos^2 i \\
\end{array}
\right).\lb{supermat}
\eqf
\Rfr{supermat} agrees with Equations~(48)~to~(53) of \citet{2003AJ....125.1580K} for $i\rightarrow \uppi/2-\delta,~\varepsilon\rightarrow\alpha$, where $\delta,~\alpha$ are the declination and right ascension, respectively.
\Rfr{supermat} is needed to work out the coefficients of Equations~(A.21)~to~(A.23) of \citet{2011CQGra..28a5009Z}
\begin{align}
\beta_\textrm{d} \lb{betad}&\doteq \qua{{\textsf{M}}^\textrm{d}}_{ij}{\hat{s}}_i {\hat{s}}_j - \qua{{\textsf{M}}^\textrm{d}}_{ij}{\hat{\ell}}_i^\textrm{d}{\hat{\ell}}_j^\textrm{d}, \\ \nonumber \\
\gamma_\textrm{d} &\doteq 2\qua{{\textsf{M}}^\textrm{d}}_{ij}{\hat{s}}_i {\hat{\ell}}_j^\textrm{d}, \\ \nonumber\\
\theta_\textrm{d} \lb{deltad}&\doteq \qua{{\textsf{M}}^\textrm{d}}_{ij}{\hat{s}}_i {\hat{s}}_j + 2\qua{{\textsf{M}}^\textrm{d}}_{ij}{\hat{\ell}}_i^\textrm{d}{\hat{\ell}}_j^\textrm{d},
\end{align}
which are the building blocks of the calculation of $\delta\tau_\textrm{prop}^{J_2}$ along with \rfrs{Ed}{Vd}.

Finally, the quadrupole-induced propagation delay due to the deflector d can be obtained by evaluating \citep{2011CQGra..28a5009Z}
\eqi \delta\tau_\textrm{prop}^{J_2} = \rp{G}{c^3}\ton{\beta_\textrm{d}{\mathcal{E}}_\textrm{d} + \gamma_\textrm{d}{\mathcal{F}}_\textrm{d} + \theta_\textrm{d}{\mathcal{V}}_\textrm{d} }\eqf
onto the unperturbed Keplerian ellipse by means of \rfrs{Ed}{Vd} and \rfrs{betad}{deltad}. The resulting explicit expression is
\eqi
\delta\tau^{J_2}_\textrm{prop} = \rp{Gm_\textrm{d} J_2^\textrm{d} R^2_\textrm{d}\ton{1+e\cos f}^2}{4c^3p^2}\mathcal{T}\ton{I,~\Omega,~u,~i,~\varepsilon},\lb{dtq}
\eqf
with
\begin{align}
\mathcal{T} \nonumber \lb{urca}&= \rp{2\ton{2+\sI\su}\sin^2 i}{\ton{1+\sI\su}^2}\grf{  \qua{ \cos 2\Omega  +\cI\sin 2\Omega\sin 2u   - \ton{1 + \cos^2 I}\cos 2\Omega \sin^2 u     }\cee  -\right. \\ \nonumber \\
\nonumber &-\left. \qua{ \cI\sin 2u +\sin 2\Omega   -2\cI\sin^2\Omega\sin 2u  -\sin 2\Omega\ton{1+\cos^2 I}\sin^2 u   }\see}  -\\ \nonumber \\
\nonumber &- 4\sin 2 i\qua{  \ton{\cO\cu - \cI\sO\su}\ce + \ton{\cu\sO + \cI\cO\su}\se    } +\\ \nonumber \\
&+ \sI\su\ton{-1 - 3 \cos 2 i}.
\end{align}
It is interesting to note that
\rfr{dtq} does not vanish for circular orbits. Furthermore, from \rfr{dtq} it turns out that there is no net quadrupolar propagation delay per cycle.

If the pulsar-SMBH in the GC is considered, the quadrupole Shapiro-type time delay is much smaller than the orbital time shift.
Indeed, by using \rfrs{dtq}{urca} for a S2-type orbital configuration, it turns out that
\begin{align}
\left.\delta\tau_\textrm{prop}^{Q_2}\right|^\textrm{max} \lb{maxQBH}& = 0.6~\mu\textrm{s}~\ton{i_\bullet = 90~\textrm{deg},~\varepsilon_\bullet = 281.8~\textrm{deg},~f = 339.8~\textrm{deg}}, \\ \nonumber \\
\left.\delta\tau_\textrm{prop}^{Q_2}\right|^\textrm{min} \lb{minQBH}& = -1.1~\mu\textrm{s}~\ton{i_\bullet = 35.8~\textrm{deg},~\varepsilon_\bullet = 182.5~\textrm{deg},~f = 349.0~\textrm{deg}}.
\end{align}
The bounds of \rfrs{maxQBH}{minQBH} should be compared with those of \rfrs{megaQ2max}{megaQ2min}, which are about four orders of magnitude larger. Values as little as those of \rfrs{maxQBH}{minQBH} should be hard to be detectable in view of the expected pulsar timing precision, even in the optimistic case of $1-10~\mu\textrm{s}$ \citep{2016ApJ...818..121P,2017IJMPD..2630001G}. If the orbital configuration of S2 is abandoned letting $I,~\Omega,~\omega,~f,~i_\bullet,~\varepsilon_\bullet$ freely vary within their full natural ranges, we get
values which can reach the $100~\mu\textrm{s}$ level
for $0\leq e\leq 0.97,~5~\textrm{yr}\leq\Pb\leq 16~\textrm{yr}$.
\section{Summary and conclusions}\lb{fine}
In order to perform sensitivity studies, designing suitable tests and reinterpreting existing data analyses in a way closer to the actual experimental practice in pulsar timing, we devised a method to analytically calculate the shifts $\Delta\delta\tau_\textrm{p}^A$ experienced by the orbital component of the time changes $\delta\tau_\textrm{p}$ of a binary pulsar p due to some perturbing post-Keplerian accelerations $A$: Schwarzschild, Lense-Thirring and mass quadrupole.
We applied it to the still  hypothetical scenario encompassing an emitting neutron star which orbits the Supermassive Black Hole in Sgr A$^\ast$; its timing precision could reach $100~\mu\textrm{s}$, or, perhaps, even $1-10~\mu\textrm{s}$ \citep{2016ApJ...818..121P,2017IJMPD..2630001G}. The main results of the present study are resumed in Table \ref{resume}. By assuming a S2-like orbital configuration and a time span as long as its orbital period, the magnitude of the post-Newtonian Schwarzschild-type gravitoelectric signature can reach $\left|\Delta\delta\tau_\textrm{p}^\textrm{GE}\right|\lesssim 10^3~\textrm{s}$. The post-Newtonian Lense-Thirring gravitomagnetic and quadrupolar effects are much smaller, amounting to at most
$\left|\Delta\delta\tau_\textrm{p}^\textrm{LT}\right|\lesssim 0.6~\textrm{s},\left|\Delta\delta\tau_\textrm{p}^{Q_2}\right|\lesssim 0.04~\textrm{s}$, depending on the orientation of the Black Hole's spin axis. Faster $\ton{\Pb = 5~\textrm{yr}}$ and more eccentric $\ton{e=0.97}$ orbits would imply net shifts per revolution $\left|\ang{\Delta\delta\tau_\textrm{p}^\textrm{GE}}\right|\lesssim 10~\textrm{Ms},~\left|\ang{\Delta\delta\tau_\textrm{p}^\textrm{LT}}\right|\lesssim 400~\textrm{s},\left|\ang{\Delta\delta\tau_\textrm{p}^{Q_2}}\right|\lesssim 10^3~\textrm{s}$ or so, depending on the other orbital parameters and the initial epoch.

Among other things, we also explicitly calculated an analytical formula for the Shapiro-like time delay $\delta\tau_\textrm{prop}$ due to the propagation of electromagnetic waves in the field of a spinning oblate body, which we applied to the aforementioned binary system. As far as the Lense-Thirring and the quadrupolar effects are concerned, the Shapiro-like time shifts $\delta\tau_\textrm{prop}$ are, in general, much smaller than the orbital ones $\Delta\delta\tau_\textrm{p}$ which, contrary to $\delta\tau_\textrm{prop}$, are cumulative. In the case of the pulsar-Sgr A$^\ast$ scenario, we have, for a S2-type orbit, that the Lense-Thirring propagation delay is as little as $\left|\delta\tau_\textrm{prop}^\textrm{LT}\right|\lesssim 0.02~\textrm{s}$, while the quadrupolar one is of the order of $\left|\delta\tau_\textrm{prop}^{Q_2}\right|\lesssim 1~\mu\textrm{s}$, both depending on the spin orientation of the Black Hole. Removing the limitation to the S2 orbital configuration yields essentially similar values for $\delta\tau_\textrm{prop}^\textrm{LT},~\delta\tau_\textrm{prop}^{Q_2}$, even for highly eccentric and faster orbits.

Finally, we remark that our approach is general enough to be extended to arbitrary orbital geometries and symmetry axis orientations of the binary's bodies, and to whatsoever disturbing accelerations. As such, it can be applied to other binary systems as probes for, say, modified models of gravity. In principle, also man-made binaries could be considered.
\section*{Acknowledgements}
I would like to thank an attentive referee for her/his precious critical remarks
\appendix
\section{Notations and definitions}\lb{appen}
Here, some basic notations and definitions used in the text are presented \citep{1991ercm.book.....B,Nobilibook87,1989racm.book.....S,2003ASSL..293.....B}
\begin{description}
\item[] $G:$ Newtonian constant of gravitation
\item[] $c:$ speed of light in vacuum
\item[] $\bds{\hat{e}}_z:$ unit vector directed along the line of sight towards the observer.
\item[] $\bds{\hat{e}}_x,~\bds{\hat{e}}_y:$ unit vectors spanning the plane of the sky.
\item[] $\mA$: mass of the body A
\item[] $\mB$: mass of the body B
\item[] $m_\textrm{p}$: mass of the pulsar p
\item[] $m_\textrm{c}$: mass of the unseen companion c
\item[] $m_\textrm{tot}\doteq \mA + \mB$: total mass of the binary
\item[] $\mu\doteq Gm_\textrm{tot}:$ gravitational parameter of the binary
\item[] $\xi\doteq\mA\mB m_\textrm{tot}^{-2}:$ dimensionless mass parameter of the binary
\item[] $S:$ magnitude of the angular momentum of any of the binary's components
\item[] $\mathcal{S}^\textrm{A/B}\doteq\ton{1+\rp{3}{4}\rp{M_\textrm{B/A}}{M_\textrm{A/B}}}S^\textrm{A/B}:$ magnitude of the scaled angular momentum of any of the binary's component
\item[] $\bds{\hat{S}}:$ unit vector of the spin axis of any of the binary's components
\item[] $i,~\varepsilon:$ spherical angles determining the spatial orientation of $\bds{\hat{S}}$; $i=90~\textrm{deg}$ implies that the latter lies in the plane of the sky
\item[] $\bds{\mathcal{S}}\doteq \sA+\sB:$ sum of the scaled angular momenta of the binary
\item[] $\chi_g:$ dimensionless angular momentum parameter of a Kerr black hole
\item[] $R:$ equatorial radius of any of the binary's components
\item[] $J_2:$ dimensionless quadrupole mass moment of any of the binary's components
\item[] $Q_2:$ dimensional quadrupole mass moment of any of the binary's components
\item[] ${\mathcal{Q}}_2^\textrm{A/B}\doteq\ton{1+\rp{M_\textrm{B/A}}{M_\textrm{A/B}}}Q_2^\textrm{A/B}:$ scaled dimensional quadrupole mass moment of any of the binary's components
\item[] $\mathbf{r}:$ relative position vector of the binary's orbit
\item[] $\mathbf{v}:$ relative velocity vector of the binary's orbit
\item[] $a:$  semimajor axis of the binary's relative orbit
\item[] $\nk \doteq \sqrt{\mu a^{-3}}:$   Keplerian mean motion
\item[] $\Pb = 2\uppi \nk^{-1}:$ Keplerian orbital period
\item[] $a_\textrm{A}=\mB M^{-1}_\textrm{tot} a:$ semimajor axis of the barycentric orbit of the binary's visible component A
\item[] $e:$  eccentricity
\item[] $p\doteq a(1-e^2):$  semilatus rectum
\item[] $I:$  inclination of the orbital plane
\item[] $\Omega:$  longitude of the ascending node
\item[] $\omega:$  argument of pericenter
\item[] $\varpi\doteq \Omega+\omega:$ longitude of pericenter
\item[] $t_p:$ time of periastron passage
\item[] $t_0:$ reference epoch
\item[] $\mathcal{M}\doteq \nk\ton{t - t_p}:$ mean anomaly
\item[] $\eta\doteq\nk\ton{t_0-t_p}:$ mean anomaly at epoch
\item[] $\lambda\doteq \varpi + \mathcal{M}:$ mean longitude
\item[] $\epsilon:$ mean longitude at epoch
\item[] $f:$  true anomaly
\item[] $u\doteq \omega + f:$  argument of latitude
\item[] $\bds{\hat{l}}\doteq\grf{\cO,~\sO,~0}:$ unit vector directed along the line of the nodes toward the ascending node
\item[] $\bds{\hat{m}}\doteq\grf{-\cI\sO,~\cI\cO,~\sI}:$ unit vector directed transversely to the line of the nodes in the orbital plane
\item[] $r:$ magnitude of the binary's relative position vector
\item[] $\uR\doteq \mathbf{r}~r^{-1}=\bds{\hat{l}}\cos u + \bds{\hat{m}}\sin u:$ radial unit vector
\item[] $\uN\doteq\grf{\sI\sO,~-\sI\cO,~ \cI}:$ unit vector of the orbital angular momentum
\item[] $\uT\doteq\uN\bds\times\uR:$ transverse unit vector
\item[] $\bds A:$ disturbing acceleration
\item[] $A_{\rho}=\bds A\bds\cdot\uR:$ radial component of $\bds A$
\item[] $A_{\sigma}=\bds A\bds\cdot\uT:$ transverse component of $\bds A$
\item[] $A_{\nu}=\bds A\bds\cdot\uN:$ normal component of $\bds A$
\item[] $\delta\tau_\textrm{p}:$ periodic variation of the time of arrivals of the pulses from the pulsar p due to its barycentric orbital motion
\end{description}
\section{Tables and Figures}
\begin{table*}
\caption{Relevant physical and orbital parameters of the S2 star and the SMBH at the GC along with their estimated uncertainties according to Table 3 of \citet{2017ApJ...837...30G}; they are referred to the epoch $2000.0$. $D_0$ is the distance to $\textrm{Sgr~A}^\ast$. The Schwarzschild radius of the SMBH is $r_g\doteq 2GM_\bullet/c^2 = 0.088~\textrm{au}$, while the linear size of the semimajor axis of S2 is $a=1,044~\textrm{au}=11,863.6~r_g$. We quote also the derived values of the SMBH's angular momentum and quadrupole mass moment calculated as \citep{1970JMP....11.2580G,1974JMP....15...46H} $S_\bullet=\chi_g M_\bullet ^2 G c^{-1},~Q_2^\bullet = - S^2_\bullet c^{-2} M^{-1}_\bullet$ due to the the \virg{no-hair} theorems \citep{1972CMaPh..25..152H,1967PhRv..164.1776I,1975PhRvL..34..905R}. The dimensionless parameter $\chi_g\leq 1$ is of the order of about $0.6$ for the SMBH in Sgr A$^\ast$ \citep{2016ApJ...818..121P}. We display also the value $f_0$ inferred from \rfr{fMt} for the true anomaly at the epoch $t_0=2003.271$ quoted in Table 5 of \citet{2017ApJ...837...30G}.}
\label{S2}
\centering
\begin{tabular}{ll}
\noalign{\smallskip}
\hline
Estimated parameter & Value \\
\hline
$M_\bullet$ & $4.28\pm \left.0.10\right|_\textrm{stat}\pm \left.0.21\right|_\textrm{sys}\times 10^6~\textrm{M}_\odot$\\
$D_0$ & $8.32\pm\left.0.07\right|_\textrm{stat}\pm \left.0.14\right|_\textrm{sys}~\textrm{kpc}$\\
$\Pb$ & $16.00\pm 0.02~\textrm{yr}$\\
$a$ & $0.1255\pm 0.0009~\textrm{arcsec}$\\
$e$  & $0.8839\pm 0.0019$\\
$I$ & $134.18\pm 0.40~\textrm{deg}$\\
$\Omega$ & $226.94\pm 0.60~\textrm{deg}$\\
$\omega$ & $65.51\pm 0.57~\textrm{deg}$\\
$t_p$ & $2002.33\pm 0.01$~\textrm{calendar~year}\\
\hline
Derived parameter & Value \\
\hline
$f_0$ & $139.72\pm 0.48$~\textrm{deg}\\
$S_\bullet$ & $\chi_g~1.61\times 10^{55}~\textrm{kg~m}^2~\textrm{s}^{-1}$ \\
$Q_2^\bullet$ & $-\chi_g^2~3.40\times 10^{56}~\textrm{kg~m}^2$\\
\hline
\end{tabular}
\end{table*}
\begin{table*}
\caption{Maximum and minimum values for the orbital and propagation time shifts $\ton{\Delta\delta\tau_\textrm{p}},~\delta\tau_\textrm{prop}$ over a full orbital revolution due to the gravitoelectric (GE), gravitomagnetic (LT) and quadrupole ($Q_2$) effects for a hypothetical pulsar-Sgr A$^\ast$ scenario. While the orbital time delay $\Delta\delta\tau_\textrm{p}$ is cumulative over the revolutions, the propagation shift $\delta\tau_\textrm{prop}$ vanishes over one orbital period. For the putative pulsar orbiting the SMBH in the GC, the orbital configuration of the main sequence S2 star was adopted.  The values of the system's physical and orbital parameters corresponding to the quoted maxima and minima are not reported here: see the text for details. Only the angular momentum $S_\bullet$ and the quadrupole $Q_2^\bullet$ of the SMBH, playing  the role of deflector d, were taken into account. The expected timing precision for a pulsar orbiting the Galactic SMBH is about $100~\mu\textrm{s}$, or, perhaps, even $1-10~\mu\textrm{s}$ \citep{2016ApJ...818..121P,2017IJMPD..2630001G}.}
\label{resume}
\centering
\begin{tabular}{ll}
\noalign{\smallskip}
\hline
$\left.\Delta\delta\tau_\textrm{p}^\textrm{GE}\right|^\textrm{max}$  & $2,520.3557~\textrm{s}$ \\
$\left.\Delta\delta\tau_\textrm{p}^\textrm{GE}\right|^\textrm{min}$ & $-6,119.2341~\textrm{s}$  \\
\hline
$\left.\Delta\delta\tau_\textrm{p}^\textrm{LT}\right|^\textrm{max}$ & $0.6054~\textrm{s}$ \\
$\left.\Delta\delta\tau_\textrm{p}^\textrm{LT}\right|^\textrm{min}$ & $-0.6054~\textrm{s}$ \\
$\left.\delta\tau_\textrm{prop}^\textrm{LT}\right|^\textrm{max}$    & $0.0195~\textrm{s}$\\
$\left.\delta\tau_\textrm{prop}^\textrm{LT}\right|^\textrm{min}$    &  $-0.0213~\textrm{s}$\\
\hline
$\left.\Delta\delta\tau_\textrm{p}^{Q_2}\right|^\textrm{max}$    & $0.0215~\textrm{s}$ \\
$\left.\Delta\delta\tau_\textrm{p}^{Q_2}\right|^\textrm{min}$    & $-0.0393~\textrm{s}$ \\
$\left.\delta\tau_\textrm{prop}^{Q_2}\right|^\textrm{max}$          & $0.6~\mu\textrm{s}$ \\
$\left.\delta\tau_\textrm{prop}^{Q_2}\right|^\textrm{min}$          & $-1.1~\mu\textrm{s}$ \\
\hline
\end{tabular}
\end{table*}

\bibliography{PXbib,IorioFupeng}{}

\begin{thebibliography}{55}
\expandafter\ifx\csname natexlab\endcsname\relax\def\natexlab#1{#1}\fi

\bibitem[{{Ang{\'e}lil}, {Saha} \& {Merritt}(2010){Ang{\'e}lil}, {Saha}, \&
  {Merritt}}]{2010ApJ...720.1303A}
{Ang{\'e}lil} R., {Saha} P., {Merritt} D., 2010, ApJ, 720, 1303

\bibitem[{{Barker} \& {O'Connell}(1975)}]{1975PhRvD..12..329B}
{Barker} B.~M., {O'Connell} R.~F., 1975, Phys. Rev. D, 12, 329

\bibitem[{{Bejger}, {Bulik} \& {Haensel}(2005){Bejger}, {Bulik}, \&
  {Haensel}}]{2005MNRAS.364..635B}
{Bejger} M., {Bulik} T., {Haensel} P., 2005, MNRAS, 364, 635

\bibitem[{{Bertotti}, {Farinella} \& {Vokrouhlick\'{y}}(2003){Bertotti},
  {Farinella}, \& {Vokrouhlick\'{y}}}]{2003ASSL..293.....B}
{Bertotti} B., {Farinella} P., {Vokrouhlick\'{y}} D., 2003, {Physics of the
  Solar System - Dynamics and Evolution, Space Physics, and Spacetime
  Structure.} Kluwer, Dordrecht

\bibitem[{{Broderick} {et~al}\mbox{.}(2009){Broderick}, {Fish}, {Doeleman}, \&
  {Loeb}}]{2009ApJ...697...45B}
{Broderick} A.~E., {Fish} V.~L., {Doeleman} S.~S., {Loeb} A., 2009, ApJ, 697,
  45

\bibitem[{{Broderick} {et~al}\mbox{.}(2011){Broderick}, {Fish}, {Doeleman}, \&
  {Loeb}}]{2011ApJ...735..110B}
{Broderick} A.~E., {Fish} V.~L., {Doeleman} S.~S., {Loeb} A., 2011, ApJ, 735,
  110

\bibitem[{{Brouwer} \& {Clemence}(1961)}]{1961mcm..book.....B}
{Brouwer} D., {Clemence} G.~M., 1961, {Methods of Celestial Mechanics}.
  Academic Press, New York

\bibitem[{{Brumberg}(1991)}]{1991ercm.book.....B}
{Brumberg} V.~A., 1991, {Essential Relativistic Celestial Mechanics}. Adam
  Hilger, Bristol

\bibitem[{{Capderou}(2005)}]{2005som..book.....C}
{Capderou} M., 2005, {Satellites: Orbits and missions}. Springer, Berlin

\bibitem[{{Casotto}(1993)}]{1993CeMDA..55..209C}
{Casotto} S., 1993, Celestial Mechanics and Dynamical Astronomy, 55, 209

\bibitem[{{Damour} \& {Deruelle}(1986)}]{DamDer86}
{Damour} T., {Deruelle} N., 1986, {Annales de l' I.H.P., section A}, 44, 263

\bibitem[{{Damour} \& {Schaefer}(1991)}]{1991PhRvL..66.2549D}
{Damour} T., {Schaefer} G., 1991, Phys. Rev. Lett., 66, 2549

\bibitem[{{Damour} \& {Schafer}(1988)}]{1988NCimB.101..127D}
{Damour} T., {Schafer} G., 1988, Nuovo Cimento B, 101, 127

\bibitem[{{Damour} \& {Taylor}(1992)}]{1992PhRvD..45.1840D}
{Damour} T., {Taylor} J.~H., 1992, Phys. Rev. D, 45, 1840

\bibitem[{{Doroshenko} \& {Kopeikin}(1995)}]{1995MNRAS.274.1029D}
{Doroshenko} O.~V., {Kopeikin} S.~M., 1995, MNRAS, 274, 1029

\bibitem[{{Dymnikova}(1986)}]{1986SvPhU..29..215D}
{Dymnikova} I.~G., 1986, Soviet Physics Uspekhi, 29, 215

\bibitem[{{Eatough} {et~al}\mbox{.}(2013){Eatough}, {Kramer}, {Klein},
  {Karuppusamy}, {Champion}, {Freire}, {Wex}, \& {Liu}}]{2013IAUS..291..382E}
{Eatough} R.~P., {Kramer} M., {Klein} B., {Karuppusamy} R., {Champion} D.~J.,
  {Freire} P.~C.~C., {Wex} N., {Liu} K., 2013, in IAU Symposium, Vol. 291,
  Neutron Stars and Pulsars: Challenges and Opportunities after 80 years, {van
  Leeuwen} J., ed., pp. 382--384

\bibitem[{{Geroch}(1970)}]{1970JMP....11.2580G}
{Geroch} R., 1970, Journal of Mathematical Physics, 11, 2580

\bibitem[{{Gillessen} {et~al}\mbox{.}(2017){Gillessen}, {Plewa}, {Eisenhauer},
  {Sari}, {Waisberg}, {Habibi}, {Pfuhl}, {George}, {Dexter}, {von Fellenberg},
  {Ott}, \& {Genzel}}]{2017ApJ...837...30G}
{Gillessen} S. {et~al.}, 2017, ApJ, 837, 30

\bibitem[{{Goddi} {et~al}\mbox{.}(2017){Goddi}, {Falcke}, {Kramer}, {Rezzolla},
  {Brinkerink}, {Bronzwaer}, {Davelaar}, {Deane}, {de Laurentis}, {Desvignes},
  {Eatough}, {Eisenhauer}, {Fraga-Encinas}, {Fromm}, {Gillessen}, {Grenzebach},
  {Issaoun}, {Jan{\ss}en}, {Konoplya}, {Krichbaum}, {Laing}, {Liu}, {Lu},
  {Mizuno}, {Moscibrodzka}, {M{\"u}ller}, {Olivares}, {Pfuhl}, {Porth},
  {Roelofs}, {Ros}, {Schuster}, {Tilanus}, {Torne}, {van Bemmel}, {van
  Langevelde}, {Wex}, {Younsi}, \& {Zhidenko}}]{2017IJMPD..2630001G}
{Goddi} C. {et~al.}, 2017, Int. J. Mod. Phys. D, 26, 1730001

\bibitem[{{Hansen}(1974)}]{1974JMP....15...46H}
{Hansen} R.~O., 1974, Journal of Mathematical Physics, 15, 46

\bibitem[{{Hawking}(1972)}]{1972CMaPh..25..152H}
{Hawking} S.~W., 1972, Communications in Mathematical Physics, 25, 152

\bibitem[{{Iorio}(2005)}]{2005A&A...433..385I}
{Iorio} L., 2005, A\&A, 433, 385

\bibitem[{{Iorio}(2007)}]{2007Ap&SS.312..331I}
{Iorio} L., 2007, Astrophys. Space Sci., 312, 331

\bibitem[{{Iorio}(2011)}]{2011PhRvD..84l4001I}
{Iorio} L., 2011, Phys. Rev. D, 84, 124001

\bibitem[{{Israel}(1967)}]{1967PhRv..164.1776I}
{Israel} W., 1967, Physical Review, 164, 1776

\bibitem[{{Johannsen}(2016)}]{2016CQGra..33k3001J}
{Johannsen} T., 2016, Classical Quant. Grav., 33, 113001

\bibitem[{{Kaspi} \& {Kramer}(2015)}]{Kaspi15}
{Kaspi} V., {Kramer} M., 2015, in {Proceedings of the 26th Solvay Conference on
  Physics on Astrophysics and Cosmology}, {Blandford} R., {Sevrin} A., eds.,
  World Scientific, Singapore, pp. 22--61

\bibitem[{{Klioner}(1991)}]{1991SvA....35..523K}
{Klioner} S.~A., 1991, Sov. Astron., 35, 523

\bibitem[{{Klioner}(2003)}]{2003AJ....125.1580K}
{Klioner} S.~A., 2003, AJ, 125, 1580

\bibitem[{{Konacki}, {Maciejewski} \& {Wolszczan}(2000){Konacki},
  {Maciejewski}, \& {Wolszczan}}]{2000ApJ...544..921K}
{Konacki} M., {Maciejewski} A.~J., {Wolszczan} A., 2000, ApJ, 544, 921

\bibitem[{{Kopeikin}(1997)}]{1997JMP....38.2587K}
{Kopeikin} S.~M., 1997, J. Math. Phys., 38, 2587

\bibitem[{{Kramer}(2016)}]{2016IJMPD..2530029K}
{Kramer} M., 2016, Int. J. Mod. Phys. D, 25, 1630029

\bibitem[{{Laarakkers} \& {Poisson}(1999)}]{1999ApJ...512..282L}
{Laarakkers} W.~G., {Poisson} E., 1999, ApJ, 512, 282

\bibitem[{{Milani}, {Nobili} \& {Farinella}(1987){Milani}, {Nobili}, \&
  {Farinella}}]{Nobilibook87}
{Milani} A., {Nobili} A., {Farinella} P., 1987, {Non-gravitational
  perturbations and satellite geodesy}. Adam Hilger, Bristol

\bibitem[{{Morrison} {et~al}\mbox{.}(2004){Morrison}, {Baumgarte}, {Shapiro},
  \& {Pandharipande}}]{2004ApJ...617L.135M}
{Morrison} I.~A., {Baumgarte} T.~W., {Shapiro} S.~L., {Pandharipande} V.~R.,
  2004, ApJL, 617, L135

\bibitem[{{Nobili}, {Milani} \& {Farinella}(1988){Nobili}, {Milani}, \&
  {Farinella}}]{1988AJ.....95..576N}
{Nobili} A.~M., {Milani} A., {Farinella} P., 1988, AJ, 95, 576

\bibitem[{{Nobili} \& {Will}(1986)}]{1986Natur.320...39N}
{Nobili} A.~M., {Will} C.~M., 1986, Nature, 320, 39

\bibitem[{{Psaltis}, {Wex} \& {Kramer}(2016){Psaltis}, {Wex}, \&
  {Kramer}}]{2016ApJ...818..121P}
{Psaltis} D., {Wex} N., {Kramer} M., 2016, ApJ, 818, 121

\bibitem[{{Robinson}(1975)}]{1975PhRvL..34..905R}
{Robinson} D.~C., 1975, Phys. Rev. Lett., 34, 905

\bibitem[{{Roy}(2005)}]{2005ormo.book.....R}
{Roy} A.~E., 2005, {Orbital Motion. Fourth Edition}. Institute of Physics
  Publishing, Bristol

\bibitem[{{Rubincam}(1977)}]{1977CeMec..15...21R}
{Rubincam} D.~P., 1977, Celestial Mechanics, 15, 21

\bibitem[{{Sahni} \& {Shtanov}(2008)}]{2008IJMPD..17..453S}
{Sahni} V., {Shtanov} Y., 2008, Int. J. Mod. Phys. D, 17, 453

\bibitem[{{Sch{\"a}fer}(2004)}]{2004GReGr..36.2223S}
{Sch{\"a}fer} G., 2004, Gen. Relat. Gravit., 36, 2223

\bibitem[{{Sch{\"a}fer}(2009)}]{2009SSRv..148...37S}
{Sch{\"a}fer} G., 2009, Space Sci. Rev., 148, 37

\bibitem[{{Soffel}(1989)}]{1989racm.book.....S}
{Soffel} M.~H., 1989, {Relativity in Astrometry, Celestial Mechanics and
  Geodesy}. Springer-Verlag; Berlin Heidelberg New York

\bibitem[{{Tapley}, {Schutz} \& {Born}(2004){Tapley}, {Schutz}, \&
  {Born}}]{2004Tapleyetal}
{Tapley} B.~D., {Schutz} B.~E., {Born} G.~H., 2004, {Statistical Orbit
  Determination}. Elsevier, Amsterdam

\bibitem[{{Thorne}(1988)}]{1988nznf.conf..573T}
{Thorne} K.~S., 1988, in Near Zero: New Frontiers of Physics, {Fairbank} J.~D.,
  {Deaver} Jr. B.~S., {Everitt} C.~W.~F., {Michelson} P.~F., eds., pp. 573--586

\bibitem[{{Wex}(2014)}]{Wex14}
{Wex} N., 2014, in {Frontiers in Relativistic Celestial Mechanics Volume 2
  Applications and Experiments}, {Kopeikin} S., ed., de Gruyter, Berlin, pp.
  39--102

\bibitem[{{Wex} \& {Kopeikin}(1999)}]{1999ApJ...514..388W}
{Wex} N., {Kopeikin} S.~M., 1999, ApJ, 514, 388

\bibitem[{{Will}(2014)}]{2014PhRvD..89d4043W}
{Will} C.~M., 2014, Phys. Rev. D, 89, 044043

\bibitem[{{Xu}(2008)}]{2008orbi.book.....X}
{Xu} G., 2008, {Orbits}. Springer, Berlin

\bibitem[{{Yu}, {Zhang} \& {Lu}(2016){Yu}, {Zhang}, \&
  {Lu}}]{2016ApJ...827..114Y}
{Yu} Q., {Zhang} F., {Lu} Y., 2016, Astrophys. J., 827, 114

\bibitem[{{Zhang}, {Lu} \& {Yu}(2014){Zhang}, {Lu}, \&
  {Yu}}]{2014ApJ...784..106Z}
{Zhang} F., {Lu} Y., {Yu} Q., 2014, ApJ, 784, 106

\bibitem[{{Zschocke} \& {Klioner}(2011)}]{2011CQGra..28a5009Z}
{Zschocke} S., {Klioner} S.~A., 2011, Classical Quant. Grav., 28, 015009

\end{thebibliography}

\end{document}